\documentclass[aps,pre,reprint,superscriptaddress]{revtex4-2}
\usepackage[centertags]{amsmath}
\usepackage{amsfonts}
\usepackage{float}
\usepackage{orcidlink}
\usepackage{CJKutf8}

\newcommand{\id}{\textrm{d}}
\let\ve=\varepsilon 
\let\oldsqrt\sqrt
\def\sqrt{\mathpalette\DHLhksqrt}
\def\DHLhksqrt#1#2{
	\setbox0=\hbox{$#1\oldsqrt{#2\,}$}\dimen0=\ht0
	\advance\dimen0-0.2\ht0
	\setbox2=\hbox{\vrule height\ht0 depth -\dimen0}
	{\box0\lower0.4pt\box2}}

\usepackage{nomencl}
\makenomenclature
\usepackage{subcaption}
\newcommand{\zvec}[1]{\vec{#1}\mkern2mu\vphantom{#1}}
\usepackage{booktabs}   

\begin{document}
\begin{CJK*}{UTF8}{gbsn} 
\date{\today}
\title{Coupling an elastic string to an active bath:\\ the emergence of inverse damping}

\affiliation{Department of Physics and Astronomy, KU Leuven}
\affiliation{School of Physics, Peking University, Beijing}

\author{Aaron Beyen \orcidlink{0000-0002-4341-7661}}
\email{Contact author: aaron.beyen@kuleuven.be}
\affiliation{Department of Physics and Astronomy, KU Leuven}

\author{Christian Maes \orcidlink{0000-0002-0188-697X}}
\affiliation{Department of Physics and Astronomy, KU Leuven}

\author{Ji-Hui Pei (裴继辉)\orcidlink{0000-0002-3466-4791}}
\affiliation{Department of Physics and Astronomy, KU Leuven}
\affiliation{School of Physics, Peking University, Beijing}

\begin{abstract}
We consider a slow elastic string with Klein-Gordon dynamics coupled to a bath of run-and-tumble particles. We derive and solve the induced Langevin-Klein-Gordon string dynamics with explicit expressions for the streaming term, friction coefficient, and noise variance. These parameters are computed exactly in a weak coupling expansion. The induced friction is a sum of two terms: one entropic, proportional to the noise variance as in the Einstein relation for a thermal equilibrium bath, and a frenetic contribution that can take both signs. The frenetic part wins for higher bath persistence, making the total friction negative, and hence creating a wave instability akin to inverse Landau damping. However, this acceleration decreases and eventually disappears when the propulsion speed of the active particles becomes much higher. Detailed simulations confirm the initial growth driven by this anti-damping.
\end{abstract}

\maketitle
\end{CJK*}

\section{Introduction}
Active particles are omnipresent in recent studies of nonequilibrium physics, \cite{TONER2005170, activematterrev1, teVrugt2025, Gompper2020, activeparticle1, activeparticle2, activematter}. 
Run-and-tumble particles (RTP) are an interesting example where, at least in one dimension, the persistent motion can be summarized in terms of a bimodal stationary velocity distribution, largely deviating from the (thermal) Maxwellian, \cite{rtp_distribution}. It is not so strange to surmise that this characteristic feature of RTP is relevant for interactions with continuous media (\textit{e.g.} strings,  membranes, waves, scalar fields). This then is the subject of the present paper: to observe the exchange and emergence of activity and persistence in the reduced dynamics of a string due to its contact with an active bath.
The string represents a one-dimensional membrane described by the Klein-Gordon equation and is ``slower'' in comparison with the swiftness of the active particles. In other words, we treat the string as the analogue of a Brownian particle \cite{Brown01091828, einsteinbrown} that is bombarded with fast active particles. \\
This setup connects with a biological context, where tissue, cell membranes are coupled to a variety of motors, driven ions, and biomolecules \cite{prost1996shape, manneville1999active, ramaswamy2000nonequilibrium, lin2006nonequilibrium, loubet2012effective, lacoste2014update, turlier2019unveiling, softcurvedwall, elasticmembraneactive, activatingmembranes}. There have also been many theoretical, numerical, and experimental studies on microscopic active particles in contact with polymers, \cite{softcurvedwall, polymerswelling, brush, reexpansionpolymer, crowedpolymer, polymerlooping, polymerswell2, buckling}.
At the same time, the Klein-Gordon equation is also a continuum version for the dynamics of lattice vibrations and, in its overdamped version, represents an Edwards-Wilkinson elastic manifold \cite{edwards}. As a possible third context, wave-particle interaction is of fundamental importance in plasma and fluids \cite{quantumplasma, cosmicray, waveparticlereview, LIRudakov_1971, bouncingwater} as well, and we are interested in its coupling to active matter; see also \cite{spinandvelocity, Hamiltonianflocks, casiulis2025geometricconditionrobotswarmcohesion} for interesting connections. Finally, we agree with \cite{diskvsmembrane} that studying the coupling of active particles with strings/membranes, as in the present work, might lead to applications in macroscopic active matter, notably in robotics, \cite{Robotics1, robots2}.

In the tradition of dynamical fluctuation theory \cite{response_theory, Baiesi2009, Baiesi2009_2, Baiesi_2013, pei2024inducedfrictionprobemoving, activefluctuationtheory,nakajima1958quantum, Itami2017, Harada2006,Maes2014SecondFDT,stef,activefluctuationtheory,MaesThomas2013,natali2020local,sakaue2012dragging,hargus2025passiveobjectschiralactive,PhysRevLett.129.038001}, and following the projection-operator method \cite{zwanzig, zwanzig2001nonequilibrium, bouchard2007morizwanzigequationstimedependentliouvillian}, we derive and solve an ``effective'' Langevin Klein-Gordon equation for the string, with explicit expressions for the streaming term, friction coefficient 
and noise variance (coloured noise).  These induced quantities can be computed exactly in a weak-coupling expansion. In that sense, our methodology differs from posing models of fluctuating strings and membranes, \cite{brownianstring, KULKARNI2023105240} (and references therein), or of nonequilibrium field theories, \cite{cates, cates2, vrugt2022derivepredictiveactivefield,kamenev, qftcritical}, where
a constant friction coefficient and (active) noise are manually added.\\

\begin{figure}[H]
    \centering
    \includegraphics[width=0.9\linewidth]{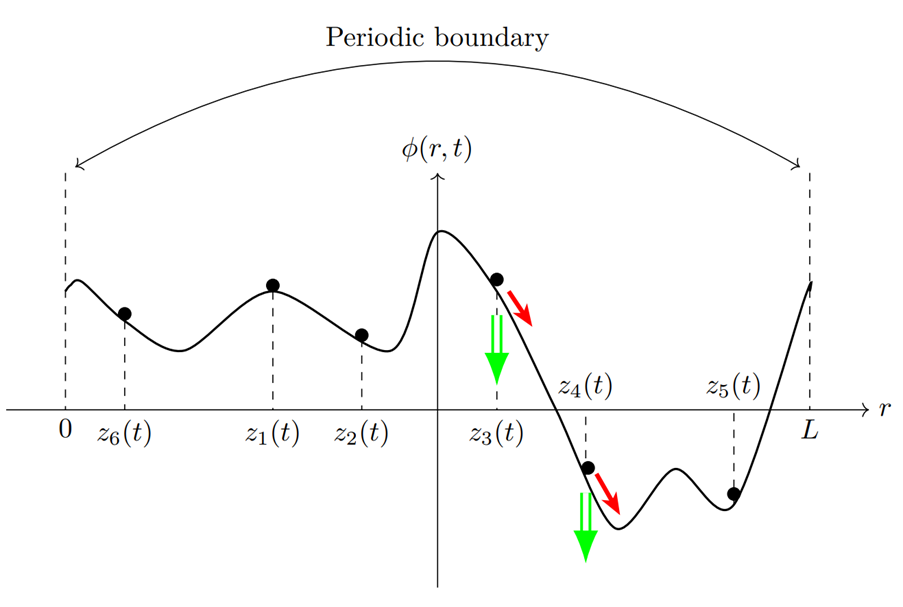}
    \caption{\raggedright A configuration of active particles coupled to the field $\phi$ on a circle. The forces on the particles and string are respectively indicated with red and green arrows: the field $\phi$ is pushed down near the particle position (green arrow), while the $z$-particles perform a gradient descent along $\phi$ (red arrow).} 
    \label{tikzpicturephi}
\end{figure}

Our main result is the emergence of negative friction in the string dynamics if the RTPs are sufficiently (but not too) persistent. It implies that the displacement variable experiences at each point an acceleration (anti-damping), resulting in the creation and propagation of many linear waves. That is reminiscent of inverse Landau damping \cite{landaudamping, quantumplasma, plasmaphysicsbook} in the study of the Vlasov-Maxwell equations when the velocity distribution of the particles is peaked at large velocities \cite{quantumplasma, plasmaphysicsbook}.
Related instabilities due to wave-particle energy transfers include
the origin of Langmuir waves \cite{langmuir} (and the ``negative mass effect'' \cite{negativemass, negativemass2}), Faraday waves \cite{waveparticleexperiment, waveparticleexperiment2} (where a flat hydrostatic surface becomes unstable by a vibrating receptacle), and plasma heating and acceleration more generally, \cite{plasmaphysicsbook, jackson_classical_1999}. Moreover, similar instabilities have been observed in the study of active Brownian particles interacting with membranes, \cite{softcurvedwall, diskvsmembrane}. \\
From numerical simulations, this anti-damping eventually halts, the amplitude saturates and the field enters a stationary regime of an elastic string becoming active by its coupling to fast run-and-tumble particles, {similar to the result in \cite{pei2025transferactivemotionmedium}. However, the complete saturation picture is not included in the present paper.
\newpage

\nomenclature{$L$}{Length of the ring}
\nomenclature{$\zeta_\phi$}{Dimensionless coupling constant}
\nomenclature{$g$}{Amplitude of the force $G(x)$}
\nomenclature{$k$}{Reduced coupling constant $k = \zeta_\phi/Y$}

\nomenclature{$\phi$}{Field amplitude}
\nomenclature{$\Pi$}{Momentum density conjugate to $\phi$}
\nomenclature{$c$}{Wave speed}
\nomenclature{$Y$}{Young's modulus of the field}
\nomenclature{$m$}{Mass density of the field}
\nomenclature{$\kappa_0$}{Spring constant (per unit length) in body force}
\nomenclature{$M$}{Reduced spring constant $M^2 = \kappa_0/Y$.}

\nomenclature{$N$}{Number of RTP particles}
\nomenclature{$z_i$}{Position of the $i$th particle, $i \in \{1,...,N \}$}
\nomenclature{$s_i$}{Spin of the $i$th particle, $i \in \{1,...,N \}$}
\nomenclature{$v_0$}{Propulsion speed RTP particles}
\nomenclature{$\alpha$}{Tumbling rate RTP particles}
\nomenclature{$\mu$}{Mobility RTP particles}

\renewcommand{\nomname}{Glossary}
\printnomenclature
\markboth{}{}

\color{black}

\section{Setup}\label{section setup}
\subsection{Equations of motion}
We consider a displacement field $\phi(r,t) \in \mathbb{R}$ (dimension of length) at time $t$ for $r\in S_L^1$, describing a dynamical string on a circle of size $L$.  Thinking about mass density $m$ and Young's modulus $Y = m c^2$ (with propagation speed $c$), we have the elasticity equations \cite{landau1elasticity, brownianstring}, 
\begin{eqnarray}\label{ela}
\frac{\partial}{\partial t}\left(m\frac{\partial \phi}{\partial t}(r,t)\right) &=& \frac{\partial \sigma}{\partial r}(r,t) + F_\phi(r,t)
\end{eqnarray}
where $\sigma(r,t) = Y\, \frac{\partial \phi}{\partial r}(r,t)$ is the stress field and $F_\phi$ is a body force (per unit length).  The body force contains two components, a confining force with spring constant (per unit length) $\kappa_o\geq 0$ plus a medium force, derived from a potential $ \zeta_\phi U_\phi(z)= \zeta_\phi\,\oint \id r  \ G(r-z) \ \phi(r)$, applied by $N$ independent active particles with positions $z_i(t) \in S^1_L$:
\begin{align}
    F_\phi(r,t) &= -\kappa_o\,\phi(r,t) - \zeta_\phi\, \sum_{i=1}^N \frac{\delta U_\phi}{\delta \phi(r,t)}(z_i(t)) \label{body force}
\end{align}
where $\zeta_\phi$ is the dimensionless coupling. The function $G$ is a smooth, periodic force per length, $G(x) = G(x + L)$, peaked around $x = 0$ with Fourier modes 
\begin{align}\label{fourier modes}
     G_n = \frac{1}{L} \oint \id x \ G(x) \ e^{- i 2 \pi n x/L}, \qquad \lim_{n \to \infty} n \ G_n = 0
\end{align}
For $G$ as a function of  $r-z_i(t)$, we need the smallest distance on the circle. The field dynamics \eqref{ela}--\eqref{body force} is then equivalent to a Klein-Gordon equation with momentum density $\Pi(r,t) =  m\frac{\partial \phi}{\partial t}(r,t) $,
 \begin{align}
 \hspace{-0.15 cm} \left(\frac{1}{c^2} \frac{\partial^2 }{\partial t^2} - \frac{\partial^2}{\partial r^2} + M^2 \right) \phi(r,t) = - k\,\sum_{i = 1}^N G(r-z_i(t))  \label{original klein gordon equation} 
  \end{align}
where $M^2 = \kappa_o/Y$,  and $k= \zeta_\phi/Y$. See also \cite{kleingordonstring, morse1953methods} for a mechanical interpretation of the Klein-Gordon equation. We take large $m$ and small $c$, but $Y>0$ remains nonzero so that the dynamics stays away from that of independent kicked harmonic oscillators.  \\

The dynamics of the active bath particles is reciprocally coupled to the string and subject to dichotomous noise,
\begin{align}
 \frac{\id z_i}{\id t}(t) &=   \mu\, f_\phi(z_i) + v_0 \,s_i(t) \label{probe dynamics}, \quad f_{\phi}(z) = - \zeta_\phi \partial_z U_{\phi}(z)
 \end{align}
with mobility $\mu$, (common) propulsion speed $v_0$, and
spins $s_i=\pm 1, i=1,\ldots,N$ that randomly flip $s_i\leftrightarrow -s_i$ independently at a rate $\alpha$. That tumbling is the only stochastic ingredient in the dynamics of $(\phi, z)$. Note that the active noise $v_0 s_i$ reduces to thermal noise $\sqrt{2 \mu  k_B T} \xi_i$ at temperature $T$ in the passive (thermal equilibrium) limit, \cite{rtp_distribution}, $v_0 \to \infty, \alpha \to \infty$ keeping the ratio $v_0^2 / \alpha = 2\mu k_B T$ fixed. This limit
inspires to define an effective inverse temperature $\beta_{\text{eff}} = 2 \alpha \mu / v_0^2$.} \\
In \eqref{probe dynamics}, the force by the field is conservative and induces a gradient flow along the field $\phi$, weighted by $G$.  The right-hand side of \eqref{original klein gordon equation} couples the string with the positions $z_i(t)$ of active particles so that an inhomogeneous force at $r = z_i(t)$ locally pushes/pulls on the string. When $\zeta_\phi G(x) > 0$ ($< 0$), it reduces (increases) the field $\phi$ and promotes its convex (concave) shape near the particle location; see Fig.~\ref{tikzpicturephi}. \\

The string-particle interaction follows  \cite{Gambassi1, Gambassi2} with potential $U_{\phi} $, and we think of the string interacting locally within the spatial extent of an active colloid, determined by the support of $G(x)$.  We prefer a smooth function on the circle that keeps track of the particle size {\it e.g.} the von Mises distribution as the circular analogue of the normal distribution, \cite{evans2000statistical},
    \begin{equation}\label{vmo}
        G(x) = \frac{g}{I_0(p)} \ \exp \left[p \cos\left(\frac{2\pi}{L}x \right)\right]
       \end{equation}
with $I_0(p)$  the modified Bessel function of the first kind and order $0$, and $p \geq  0$. Its maximum lies at $x = 0$ and the ratio of maximum to minimum values equals $e^{2p}$. 
The difference between $\pm v_0$ and $\mu \,f_\phi(z)$ plays an important role in \eqref{probe dynamics}. For the particles to move around the circle, we assume that, at least initially, $v_0 > \mu\,|f_\phi(z_i)|$. 
 Of course, since $\phi$ has a dynamics on its own, this assumption can eventually break down; see Section \ref{section remarks and outlook}, after which the particles get stuck and push in one of the potential wells of $U_\phi$ created by $\phi$.

\subsection{Time scale separation}

Integrating out the active particles requires a time scale separation in the joint dynamics. We consider the generators of the time-evolution, as given by the forward generators $\mathcal{L}^{\dagger}_{z,s}, \mathcal{L}^{\dagger}_{\phi}, \mathcal{L}^{\dagger}_{\Pi}$ for the active particle $z$ (with spin $s$), the scalar field $\phi$ and its momentum density $\Pi$,  and we select their relevant time scales; see {the Supplemental Material (SM) \cite{supp}. By dimensional analysis we arrive at} the total generator $\mathcal{L}^{\dagger} = \mathcal{L}^{\dagger}_{z,s} + \ve_\phi \mathcal{L}^{\dagger}_{\phi} + \ve_\Pi \mathcal{L}^{\dagger}_{\Pi}$, where
\begin{align}\label{first epsilon}
    \ve_\phi = \frac{\Pi_c L}{m v_0 \phi_c} \ll 1 , \qquad \ve_\Pi  = \frac{m c^2 \phi_c}{L v_0 \Pi_c} =  \frac{Y  \phi_c}{L v_0 \Pi_c} \ll 1
\end{align}
with typical field and momentum scales $\phi_c, \Pi_c$, \textit{e.g.}, the initial amplitude and momentum.
Assuming the field is initially uncoupled from the particles and using the typical time scale $L/c$ of the (uncoupled) Klein-Gordon equation on the ring with propagation speed $c$, one naturally takes $\Pi_c = m \phi_0  c/L$ and obtains
\begin{align}\label{concr}
   \ve_\Pi & = \ve_\phi = \ve  = \frac{c}{v_0} \ll 1 
\end{align}
That makes the numbers \eqref{first epsilon} more concrete:
the time scale $L/c$ should be much slower than that of the active particles, $L/v_0$, for large enough $v_0$ (at fixed tumbling rate $\alpha$). In other words, our assumption of time scale separation is that $\sqrt{Y/m}\ll v_0$, 
as for a heavy string.\\ 
When integrating out the active particles,  we keep track of the terms to order $\ve^2$ and focus on the behavior of the field at time scales $t \sim O(\ve^{-2})$,  \textit{i.e.} the Markov approximation will be exact in the limit $\ve \to 0, t \to \infty$ keeping $\ve^2 t$ constant. Moreover, such time scale separation typically also requires a weak coupling $\zeta_\phi \ll 1$ in \eqref{body force}.

\section{Reduced dynamics}\label{section main result}
Following the projection-operator formalism \cite{supp, zwanzig2001nonequilibrium, bouchard2007morizwanzigequationstimedependentliouvillian}, the resulting dynamics for the field is a Markov process described by the Langevin--Klein-Gordon equation, see SM \cite{supp},
\begin{align}
&  \frac{1}{c^2} \frac{\partial^2 \phi}{\partial t^2}(r,t)   
     = \frac{\partial^2 \phi}{\partial r^2}(r,t) - M^2 \phi(r,t) + S_\phi(r) \label{langevin eq phi} \\
    & -\oint \id u \ \nu (r, u, [\phi]) \ \frac{\partial \phi}{\partial t}(u,t)   +\oint \id u \ \Gamma (r,u, [\phi]) \ \xi(u,t) \notag
\end{align}
The white noise $\xi(r,t)$ satisfies     $\langle \xi(r,t) \rangle_\xi = 0$ and
\begin{equation}\label{white noise}
 \langle \xi(r,t) \, \xi(u,t') \rangle_\xi = \sum_{\ell = - \infty}^{\infty} \delta(r-u- \ell L) \ \delta(t-t') 
\end{equation}
For the other terms, we introduce the pinned (Born-Oppenheimer) ensemble with distribution $\rho_\phi(z,s)$. It represents the stationary distribution for the RTP satisfying $\mathcal{L}_{z,s}^{\dagger} \rho_\phi(z, s) = 0$ at pinned/fixed profile $\phi(r,t)$. 
The RTPs are moving so fast compared to the field $\phi$, that the latter is essentially fixed at $\phi = \phi(r,t)$ for the particles (without backreaction), similar to the Born-Oppenheimer approximation, \cite{Born1927, szabo1996modern}. The average $\langle \cdot \rangle_{\phi}^{\text{BO}}$ and  covariance $\langle \cdot \ ; \ \cdot \rangle_{\phi}^{\text{BO}}$ are evaluated with respect to this stationary distribution.
\begin{equation}
      \left\langle  f \right\rangle_{\phi}^{\text{BO}}   = \sum_{\vec{s}}\oint \id \vec{z}  \ f(z, s) \  \rho_\phi(z, s)
\end{equation}
This distribution factorizes due to the independence of the $z$-particles $\rho_\phi(z, s) = \prod_{i = 1}^N \rho_{\phi}^i(z_i, s_i)$,
\begin{equation}
  \sum_{s_i = \pm 1} \oint \id z_i \ \rho_\phi^i(z_i, s_i) = 1
\end{equation}
and its solution can be found in Section III of the SM \cite{supp}.\\

The term $S_\phi(r)$ in \eqref{langevin eq phi} is the induced quasistatic force (streaming term), $S_\phi(r) = - k \sum_i^N \big\langle  G(r - z_i(t))\big\rangle_{\phi}^{\text{BO}}$ and is of order $O(\ve^1)$.
The other terms contribute at order $O(\ve^2)$. The functional $\nu$ is the friction coefficient per unit length (as it is multiplied by the field velocity $\partial \phi/\partial t$) and given by the covariance
\begin{align}
       &\nu(r, u, [\phi]) =  \label{reduced form nu}\\
       &- k \sum_i^N \int_0^{\infty} \id \tau \ \left \langle G(r - z_i(t+\tau)) \ ;  \ \frac{\delta \log \rho_\phi}{\delta \phi(u,t)} (z, s)   \right \rangle_\phi^{\text{BO}} \notag
\end{align}
Finally, the functional $\Gamma(r,u, [\phi])$ is the ``square root'' of the noise amplitude defined through the relation $2 B(r,u, [\phi]) = \oint  \id q \ \Gamma (r,q, [\phi]) \ \Gamma (u, q, [\phi]) $ with
\begin{align}
& B (r,u, [\phi]) =\label{reduced form D}  \\
&
k^2\, \sum_i^N \sum_j^N \int_0^{\infty} \id \tau \ \Big \langle  G(r - z_i(t+\tau)) \ ; \  G(u - z_j(t)) \Big \rangle_{\phi}^{\text{BO}} \notag
\end{align}

\subsection{Streaming term}
For the streaming term  $S_\phi(r) $ in \eqref{langevin eq phi} we need the probability density 
$p_\phi^i(z_i) = \rho_\phi^i(z_i, s_i) + \rho_\phi^i(z_i, -s_i)$ that the probe is at location $z_i$. The complete solutions of $\rho_\phi$ and $p_\phi$ are found in Section III of the SM \cite{supp}. Up to order $\zeta_\phi^2$ in the coupling constant, we find
\begin{eqnarray}
  &&  S_\phi(r) = - \zeta_\phi^2\oint \id u \  \mathcal{M}(r-u) \  \phi(u,t) - \frac{\zeta_\phi}{Y} \frac{N}{L} \oint \id x \,G(x) \nonumber  \\
    &&  \mathcal{M}(r) 
 = -\frac{N 2 \alpha \mu}{Y L v_0^2} \oint \id x \left( G(r+x) - \frac{1}{L} \oint \id y \ G(y) \right) G(x) \notag  
\end{eqnarray}
In other words, the streaming term resets the Klein-Gordon equation with a source term $\frac{\zeta_\phi}{Y} \frac{N}{L} \oint \id x \,G(x)$ which simply sets a new reference height for the field, and, at order $\zeta_\phi^2$, the kernel $\mathcal{M}(r)$ appears resulting in an effective mass $M_{\text{eff},n}$ for each field mode $\phi_n$ in Fourier space (akin to \eqref{fourier modes}), see SM \cite{supp}, 
for $n\neq 0$,
\begin{align}\label{M_eff} 
    M_{\text{eff},n} & = M \left( 1 - \zeta_\phi^2  \frac{N}{L} \,\frac {\beta_{\text{eff}}}{2\kappa_0}\, (L|G_n|)^2 \right)
\end{align}
with $\beta_{\text{eff}} = 2 \alpha \mu/v_0^2$ and $M_{\text{eff},0} = M = \lim_{n \to \infty}  M_{\text{eff},n}$.\\
Therefore, the confinement is slightly reduced.  Since $G(x)$ is a smooth and bounded function, $G_n$ decays with growing $n$
so that the confinement of large modes remains largely unaffected by the bath.  The $n = 0$ mode (= the average $\phi$) dynamics can be solved exactly and does not contain additional confinement, friction, or noise. It only contains a constant downward shift.

\subsection{Friction and noise}
\subsubsection{Decomposing the friction}
From the SM \cite{supp}, we learn the stationary distribution $\rho_\phi$ in the weak coupling limit to linear order in $\zeta_\phi$.  That yields the friction coefficient \eqref{reduced form nu}, which can be decomposed as
\begin{align}\label{friction small coupling limit}
&\nu  =  \left(\beta_{\text{eff}}  B   + \gamma  \right) Y  + O(\zeta_\phi^3) \\ 
     & \gamma (r, u, [\phi]) =- k^2 \frac{\mu}{v_0} \sum_{i,j}^N \int_0^\infty \id \tau \  \cdot \label{def gamma} \\
     & \qquad \qquad  \left\langle G(r - z_i(t+\tau)) \ ; \  s_j \  \partial_{z_j} G(u-z_j(t))   \right \rangle_\phi^{\text{BO}} \nonumber
\end{align}
with $\beta_{\text{eff}} = 2 \alpha \mu / v_0^2$. The friction in \eqref{friction small coupling limit} can thus be decomposed as a term proportional to the noise covariance $\nu_{\text{eq}} = \beta_{\text{eff}} B Y$ -- representing the standard fluctuation-dissipation relation of the second kind (FDRII) at effective temperature $\beta_{\text{eff}}$ -- and an additional active contribution $\nu_{\text{active}} = \gamma Y$ that vanishes in the passive limit, $v_0 \to \infty, \quad v_0^2 / \alpha \to 2\mu k_B T$, when we effectively deal with a thermal bath at temperature $T$. Hence, the standard FDRII is already violated in the weakly coupled (but strongly) nonequilibrium regime, and we will see in the next subsection that the (total) friction may become negative. 
The appearance of $Y$ here stems from our referring to the Langevin-Klein--Gordon equation \eqref{original klein gordon equation}, \eqref{langevin eq phi}.
In the original formulation \eqref{ela}, we get $\tilde{\nu} = \nu Y,   \tilde{\gamma} = \gamma Y^2, \tilde{B} = B Y^2$, and the friction \eqref{friction small coupling limit} becomes  $\tilde{\nu}  = \beta_{\text{eff}}\, \tilde{B}   + \tilde{\gamma}$; see SM \cite{supp}.\\  
Alternatively, following \cite{stef,pei2024inducedfrictionprobemoving,frenesy},
it is useful to 
write the friction as the sum $\nu = \nu_\text{ent} + \nu_\text{fren}$
of entropic and frenetic parts, satisfying
\begin{equation}\label{entropic frenetic split}
    \nu_{\text{ent}} =  \frac{\beta_{\text{eff}}}{2} BY, \quad  
        \nu_{\text{fren}} =  \frac{\beta_{\text{eff}}}{2} BY  + \gamma Y
\end{equation}
That split gives a meaningful decomposition in a large family of nonequilibrium systems, in essence going back to the natural division of the path-space action into a time-antisymmetric/entropic and a time-symmetric/frenetic component. These components lead to two distinct contributions in the response and indeed, the two components in the friction \eqref{entropic frenetic split}. The entropic  part $ \nu_{\text{ent}}$ still satisfies an Einstein relation at effective inverse temperature $\beta_{\text{eff}}$, while the frenetic part $\nu_{\text{fren}}$ adds an extra piece with $\gamma $ that vanishes in the passive limit, $v_0 \to \infty, \quad v_0^2/\alpha \to 2\mu k_B T$.

\vspace{-0.5 cm}

\subsubsection{Working per mode}
\begin{figure}[ht]
    \centering
    \hspace*{-0.5cm}
    \includegraphics[width=1.1\linewidth]{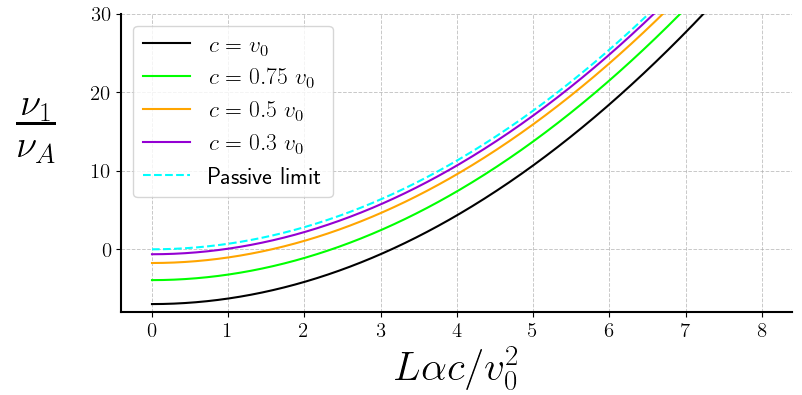}
    \caption{\raggedright Showing the dimensionless friction \eqref{nu fourier expansion} as function of the tumbling rate $\nu_1/\nu_A$ {\it vs} $L \alpha c/ v_0^2 $ with constant $\nu_A = k^2 g^2 \mu  m$ for the von Mises distribution \eqref{vmo} at different $c/v_0$ with $p = 2$,$N = 10$. The passive limit takes $v_0 \to \infty, \quad v_0^2/\alpha \to 2\mu k_B T$, $v_0 \to \infty$. The $x-$axis variable is constructed so that it is dimensionless and well-defined in the passive limit.}
    \label{nu_n vs beta}
\end{figure}

The time-dependent transition probabilities can be computed perturbatively; see SM \cite{supp}, yielding the friction $\nu $, \eqref{friction small coupling limit}, the noise coefficient $B $, \eqref{reduced form D}, and $\gamma$ from \eqref{def gamma},  all to order $\zeta_\phi^2$. They can be represented with a Fourier cosine series, \textit{e.g.} $\nu(r,u) =\nu_0 + 2 \sum_{n = 1}^\infty \nu_n \cos \left( \frac{2 \pi n}{L} (r-u) \right) $ with coefficients, for $n \neq 0$,
 \begin{align}
     \gamma_n &= - \ k^2\frac{N\mu |G_n|^2}{v_0^2}, \qquad      B_n = k^2 \frac{N L^2\alpha |G_n|^2}{2\pi^2 v_0^2 n^2} \label{B fourier expansion} \\
     \nu_n & = k^2 \frac{ NY \mu |G_n|^2}{v_0^2}  \ \left(\frac{L^2\alpha^2}{\pi^2 n^2v_0^2} - 1\right) \label{nu fourier expansion} 
 \end{align}
and $\nu_0 = B_0 = \gamma_0 = 0$.
 The functions $\nu, B, \gamma$ do not depend on $\phi$ to order $\zeta_\phi^2$. Moreover, $B $ is symmetric, $B (r,u) = B (u,r)$ and a positive-definite operator since
\begin{align*}
    & \frac{1}{L^2} \oint \id r \ \id u \ \chi(r) \  B (r,u) \ \chi(u) \\
    & = \  k^2 \frac{N L^2 \alpha}{\pi^2 v_0^2} \sum_{n = 1}^{\infty} \frac{|G_n|^2}{n^2} |\chi_n|^2 \geq 0 
\end{align*}
for all real and smooth $L$-periodic test functions with Fourier modes $\chi_n$, (as in \eqref{fourier modes}).
The ``square-root'' of $2B$ is the positive-definite, symmetric operator $\Gamma(r,u) = \Gamma_0 + 2 \sum_{n = 1} \Gamma_n \ \cos\left(\frac{2 \pi n}{L}(r-u) \right) $, with $\Gamma_0 = 0$ and for $n \neq 0,$
\begin{align}\label{Gamma_n}
\Gamma_n = k \frac{\sqrt{N L \alpha}  |G_n|}{\pi v_0 |n|} 
\end{align}
Furthermore,  $\gamma (r,u)$ in \eqref{def gamma} is always negative-definite,
\begin{align*}
   &\frac{1}{L^2} \oint \id r \  \id u \ \chi(r) \  \gamma (r,u) \ \chi(u)  \\
   & = - k^2\frac{2N\mu}{v_0^2} \sum_{n = 1}^{\infty} |G_n|^2 \ |\chi_n|^2 \leq 0
\end{align*}
The combined friction $\nu$ at the leading order then satisfies
\begin{align}\label{definitness of nu}
   &\frac{1}{L^2} \oint \id r \ \id u \ \chi(r) \  \nu (r,u) \ \chi(u) \nonumber \\
     & = \   k^2 \frac{2 N Y\mu}{v_0^2} \sum_{n = 1}^{\infty} |G_n|^2 \left(\frac{ L^2\alpha^2}{\pi^2 n^2 v_0^2} - 1\right) \ |\chi_n|^2 
  \end{align}
Thus, as can also be seen in Fig. \ref{nu_n vs beta}, the friction is negative-definite when $ \alpha < \alpha_c =  \ \pi v_0/L $. That result implies an instability in the string dynamics when the tumbling rate $\alpha$ is sufficiently small compared to the propulsion frequency $v_0/L$. However, the negativity is not monotone in the persistence; for very large $v_0$, the friction tends to zero again (at fixed tumbling rate $\alpha$). In the passive limit, $\nu_n \geq 0$. \\
Finally, shown in the SM \cite{supp},
a constant friction term $\nu_c \frac{\partial \phi}{\partial t}(r,t)$ in \eqref{langevin eq phi} (as in  \cite{brownianstring}) is not
allowed as induced friction.
\vspace{-0.5 cm}
\section{Langevin--Klein-Gordon equation}\label{section Langevin klein gordon eq}
\begin{figure}[H]
\centering
\begin{subfigure}[b]{0.45\textwidth}
	\hspace*{-0.5cm}
  \includegraphics[width=1\linewidth]{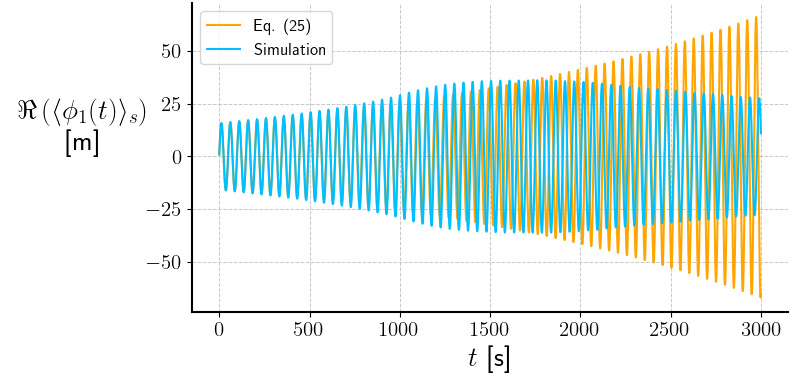}
  \caption{$\alpha < \alpha_c = \pi v_0/L$}
  \label{real part average phi1 rtp negative friction} 
\end{subfigure}
\medskip
\begin{subfigure}[b]{0.45\textwidth}
	\hspace*{-0.5cm}
  \includegraphics[width=1\linewidth]{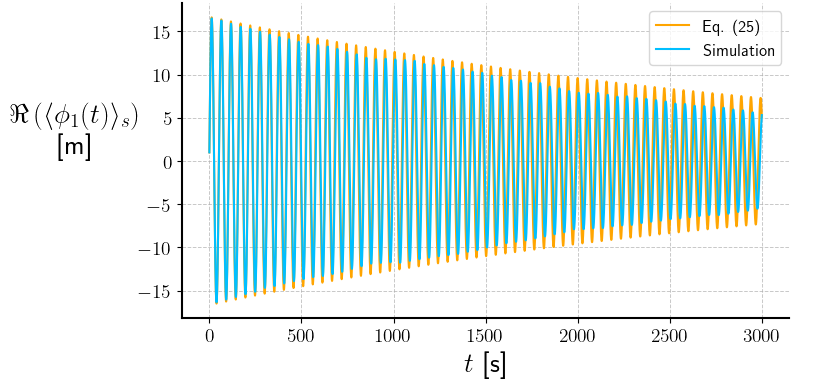}
  \caption{$\alpha > \alpha_c = \pi v_0/L$}
  \label{real part average phi1 rtp positive friction}
\end{subfigure}
\caption[Two numerical solutions]{Average of $\Re\{\phi_1(t)\}$ over 100 noise/spin realisations versus time. We show both (a)  the negative and (b) the positive friction regimes. In (a), at short times, $t < 1000$ s, the simulation result and Eq. \eqref{average phi n} overlap almost exactly and exhibit exponential growth due to the negative friction effect. At later times $t > 1500 $ s, the simulation amplitude saturates, while \eqref{average phi n} keeps growing. For (b), both the simulation result and \eqref{average phi n} decay exponentially. Other figures and the parameter values (Table I) are available in the SM \cite{supp}.}
\label{average phi1 rtp negative friction}
\end{figure}

Here we consider the stochastic behavior of the field $\phi$, induced by the active probes, as encoded in the Langevin Klein-Gordon equation \eqref{langevin eq phi}. In Fourier space \eqref{fourier modes},
\begin{align}\label{eq general potential fourier space}
    \left(\frac{\id^2}{\id t^2} + \nu_n L c^2 \frac{\id}{\id t} + \omega_n^2 \right) \phi_n(t) &= - k  NG_0 c^2 \,\delta_{n,0} \\
    & \quad + \Gamma_{n} L c^2  \xi_n(t) \nonumber 
\end{align}
 with $\nu_n, \Gamma_n$ from \eqref{nu fourier expansion}--\eqref{Gamma_n}, $\langle \xi_n(t) \rangle_\xi = 0$ and
 \begin{align}
   &\omega_n = c\sqrt{\frac{4  \pi^2 n^2}{L^2} + M_{\text{eff},n}^2}, \quad  
     \langle \xi_n(t) \ \xi_m^*(t') \rangle_\xi = \frac{\delta_{n,m}}{L} \delta(t-t') \nonumber
\end{align} 
Hence, each mode is a damped harmonic oscillator with friction coefficient $ \nu_{n}L c^2$ and (undamped)  oscillation frequency $\omega_n$ subject to a constant downward push $-k  NG_0 c^2 \,\delta_{n,0}$ and fluctuating source term $\Gamma_{n} L c^2  \xi_n(t)$.  The zero-mode experiences no friction $\nu_0 = 0$,  and the friction coefficient is negative for sufficiently large $n$ ($L \alpha  \leq \pi v_0 |n|$) unless we take the passive (thermal equilibrium) limit $v_0 \to \infty, \quad  v_0^2/\alpha \to 2\mu k_B T$. \\
The solution to \eqref{eq general potential fourier space} is easily obtained; see also \cite{brownianstring}.  Assuming mode $n$ to be underdamped, $\nu_n L c^2 < 2 \omega_n$, the homogeneous solution oscillates with a reduced frequency $\Omega_n$ (compared to the undamped case) given by,
\begin{align}\label{def omega_n}
    &\Omega_n^2 = \omega_n^2 - \frac{\nu_{n}^2 L^2}{4}  c^4
\end{align}
For $n \neq 0$, the modes satisfy
\begin{align}\label{average phi n}
   \langle \phi_n(t) \rangle_\xi  & =
 e^{- \frac{\nu_n L}{2} c^2 t} \big[ \phi_n(0) \cos\left(\Omega_n t \right)  \\
  &\qquad + \frac{\nu_n L c^2 \phi_n(0) + 2 \phi'_n(0)}{2 \Omega_n} \sin\left(\Omega_n t \right) \big] \notag
\end{align}
  and
  \begin{align}
  \hspace{-0.1 cm}  \text{Var}\left(\phi_n(t) \right)_\xi 
    &=  \frac{c^4  |\Gamma_{n}|^2 L}{\Omega_n^2}\int_0^{t} \id \tau  \  e^{- \nu_n L c^2(t-\tau)} \sin^2\left(\Omega_n (t-\tau) \right) \nonumber 
 \end{align}
This result is in agreement with \cite{brownianstring} except that we have a different friction coefficient $\nu_n$ for each mode  $n$, which can be negative. We thus recognize two possibilities for $\text{Var}\left(\phi_n(t) \right)_\xi$ (with $n \neq 0$) as $t \to \infty$:
\begin{itemize}
    \item If $\nu_{n} > 0$ (when in equilibrium or possibly for small $n$), then the integral converges, leading to a constant variance
    \begin{align}\label{fluctuations phi n at late times}
       \lim_{t \to \infty} \text{Var}\left(\phi_n(t) \right)_\xi = \frac{|\Gamma_{n}|^2 c^2}{2 \nu_{n} \omega_n^2}
    \end{align}
and the membrane is fluctuating. This result can be understood by interpreting \eqref{eq general potential fourier space} as a standard Langevin equation for two particles $\phi_n^{\text{Re}}, \phi_n^{\text{Im}}$ with ``mass'' $m \to 1/c^2$, friction coefficient $\gamma \to \nu_n L c^2$, spring constant $k \to \omega_n^2/c^2$ and effective temperature $k_B T_{\text{eff}} = |\Gamma_n|^2/(4 \nu_n)$ for which $\lim_{t \to \infty} \text{Var}\left(\phi_n(t) \right)_\xi = 2 k_B T_{\text{eff}}/k$. In equilibrium, this becomes
     \begin{align*}
        \lim_{t \to \infty}  \text{Var}\left(\phi_n(t) \right)_\xi^{\text{eq}} &=\frac{k_B T}{2L Y} \left(\frac{4 \pi^2 n^2}{L^2} + M_{\text{eff},n, \text{eq}}^2 \right)^{-1}
    \end{align*}
    which decays with $n$ and $L$.  
        \item  If $\nu_{n} < 0$ (for large $n$) then the integrand grows exponentially, indicating an instability which is easily seen in simulations of \eqref{eq general potential fourier space} as well.
\end{itemize}
To demonstrate the appearance of negative friction, we have simulated the coupled equations \eqref{original klein gordon equation}--\eqref{probe dynamics} numerically for a simple setup with $G(x) = g \left(1 + \cos \left(\frac{2 \pi}{L} x \right) \right)$ such that only the mode $\phi_1$ couples to the active particles; see Section VII in the SM \cite{supp} for more details. We compare the simulations to \eqref{eq general potential fourier space} (and its consequences like \textit{e.g.} \eqref{average phi n}). The code is available at \cite{numericalsimulation}. Most importantly, by changing the tumbling rate $\alpha$, we observe the transition from positive to negative friction, as depicted in Fig. \ref{average phi1 rtp negative friction} and theoretically predicted.

\section{Additional remarks and outlook}\label{section remarks and outlook}
\subsection{Saturation}\label{section saturation}
\begin{figure}[H]
    \centering
    \hspace*{-0.5cm}
    \includegraphics[width=1.05\linewidth]{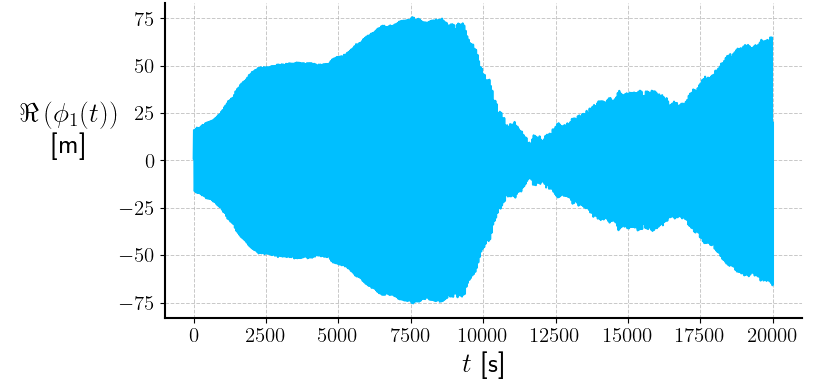}
    \caption{Individual $\Re\{\phi_1\}$ trajectory (no average) {\it vs} time, simulated over a long time interval. On this timescale, one cannot distinguish the individual oscillations, but we clearly see the negative friction effect due to the initial growth. Eventually, the amplitude saturates and shows pulsations of growth and decay.}
    \label{individual trials rtp negative friction long time}
\end{figure}

A direct follow-up question from the analysis of the variance concerns the nonlinear regime where the displacements grow quickly due to the negative linear friction, which the confining mass-term $M^2$ cannot stop. At some point, the height and local gradients of $\phi$ become too large, the active particles get stuck in the potential wells of the field, and the persistent active current around the circle halts. Then, our assumptions about the separation of time scales start to fail, and we expect a (nonlinear) saturation regime for the amplitude to appear, as in Fig. \ref{real part average phi1 rtp negative friction}. When the particles eventually escape the potential wells of $\phi$, an active particle current reappears, which again feeds the acceleration/growth in $\phi$. Qualitatively, waves are continuously created and extinguished, yielding a pulsating displacement pattern, as in Fig. \ref{individual trials rtp negative friction long time}.
Additionally, we may recall the shape transition in the steady behavior of confined run-and-tumble particles at a critical value of the tumbling rate $\alpha = \alpha_c  = \pi v_0/L$. If $\alpha > \alpha_c$, the distribution is maximal at the minima of the potential (as expected), while for $\alpha < \alpha_c$, the distribution is minimal at the extrema of the potential, where a so-called edge state \cite{rtp_distribution, faezehactivep} is created; a behavior unseen in equilibrium. One can imagine that a stable saturation regime may originate from the pushing of the active particles at the slopes (and not the minima) of the field. \\ 
For a particle probe in an active bath, as here, we know saturation occurs and the activity is inherited by the probe, \cite{pei2025transferactivemotionmedium}. The corresponding regime for waves would constitute the derivation of an active field theory and would model fluctuating surfaces that are constantly agitated much the same way as water striders perturb a still pond, \cite{besse:tel-04140962}.

\subsection{Inverse Landau-like damping}
To connect with the theory of wave dynamics, we give the homogeneous part of the solution to \eqref{eq general potential fourier space} in the SM \cite{supp}. It is of the form $  \phi_{h,n}(t) = A_{n} e^{iW^{+}_{n} t} + B_{n} e^{i W^{-}_{n} t}$ where $W_{n}^\pm$ satisfy the dispersion relations
\begin{align}\label{W}
     W_n^{\pm}& =  i \ \frac{\nu_{n} L}{2} c^2 \pm \Omega_n
\end{align}
where the imaginary part induces a positive or negative friction depending on the sign of $\nu_n$.
The $n = 0$ mode oscillates without friction, the $n \sim  O(1)$ modes experience positive or negative friction (depending on the ratio $L \alpha/(\pi v_0)$), while large modes $n \gg 1$ experience negative friction, vanishing with $n \to \infty$. That behavior is analogous to the inverse Landau-damping studied using the Vlasov-Maxwell equations, \cite{plasmaphysicsbook, quantumplasma}, where an electromagnetic wave interacting with charged particles gets damped (or antidamped) depending on the velocity distribution of the particles. As mentioned before, our wave is coupled to particles with a bimodal velocity distribution, which, in the analysis of the Vlasov equation, is known to produce a negative drag. The imaginary part of \eqref{W} vanishes in the limit $v_0 \to \infty$, implying that the active particles cannot move too fast to cause substantial negative friction. That resonance phenomenon agrees with the physics of Landau damping, where only the particles with velocity close to the phase velocity of the wave contribute to the damping (resonant effect). \\
This result aligns with other connections between active matter and charged particle systems proposed in the past, \cite{spinandvelocity, Hamiltonianflocks, casiulis2025geometricconditionrobotswarmcohesion}.

\section{Conclusions}
The interaction between continuous media and particles is a central topic in much of modern physics, and this paper specifically addresses the transfer of persistence and activity between particles and waves. Indeed, setting up a nonequilibrium dynamics of continuous (field) degrees of freedom requires understanding how that arises from coupling with active matter degrees of freedom. Within that program, we have studied a system of fast-moving, overdamped, run-and-tumble particles moving on and interacting with a slower string modeled as a scalar Klein-Gordon field. Using time scale separation and weak coupling, we have derived an effective fluctuation dynamics for the field after integrating out the active bath. \\
Akin to Landau (inverse) damping, the particles induce friction on the scalar field given by an explicit time correlation for bath observables. Depending on the level of activity and persistence of the active particles (and their velocity distribution), this friction can be negative, leading to instability. This emergence of negative (linear) friction for an elastic string extends previous results where the probe is a slow inertial particle in an active medium, \cite{pei2024inducedfrictionprobemoving, pei2025transferactivemotionmedium, kim, tailleur, caprini}, except that the acceleration (creating transverse waves) is orthogonal to the active motion.

\begin{acknowledgments}
AB is supported by the Research Foundation - Flanders (FWO) doctoral fellowship 1152725N, and JP by the
 China Scholarship Council, No. 202306010398.
\end{acknowledgments}

\nocite{Kampen1992, goldenfeld2018lectures, Bijnens_2021, Paoluzzi_2024, dichotomousnoise, mathmethods, integraltransformtables, multivariatefokkerplanck, fokkerplanckbook, rungekutta, Bena_2003, rtp_space_dependent_speed}

\bibliographystyle{unsrt}
\bibliography{bibactive.bib}

\twocolumngrid
\clearpage
\onecolumngrid 
\begin{CJK*}{UTF8}{gbsn}

\begin{center}
{\LARGE \textbf{Supplemental Material}}\\[1ex]
{\large \textbf{Coupling an elastic string to an active bath:\\ the emergence of inverse damping}}\\[2ex]

\normalfont
Aaron Beyen \orcidlink{0000-0002-4341-7661}$^{1,*}$, Christian Maes \orcidlink{0000-0002-0188-697X}$^1$ and Ji-Hui Pei (裴继辉) \orcidlink{0000-0002-3466-4791}$^{1,2}$\\
{\small $^1$\textit{Department of Physics and Astronomy, KU Leuven}} \\
$^2$\textit{School of Physics, Peking University, Beijing}\\
(Dated: \today)
\end{center}

\vspace{1em}

\end{CJK*}

\setcounter{section}{0}
\setcounter{equation}{0}

\section{Fokker-Planck equation combined system}
The dynamics for the combined system $(\phi,z)$ is given in Eq. (3)-(4) of the manuscript. Writing them out explicitly yields the form
\begin{align}
 &\left(\frac{1}{c^2} \frac{\partial^2 }{\partial t^2} - \frac{\partial^2}{\partial r^2} + M^2 \right) \phi(r,t) = - k\,\sum_{i = 1}^N G(r-z_i(t))  \label{original klein gordon equation supp} \\
  &\frac{\id z_i}{\id t}(t) =   -\mu \zeta_\phi \int \ \id r \  \partial_zG(r- z_i(t)) \ \phi(r,t)  + v_0 \,s_i(t) \label{original z dynamics}
  \end{align}
The Fokker-Planck equation is then given by, \cite{Kampen1992, goldenfeld2018lectures, qftcritical, rtp_distribution},
\begin{align}
    \frac{\partial \rho_{\text{tot}}}{\partial t}&([\phi,\Pi],z, s, t) = \mathcal{L}^{\dagger} \rho_{\text{tot}}([\phi, \Pi],z, s, t) \nonumber \\
     =& -  \sum_{i}^N \left[  \frac{\partial}{\partial z_i} \left( \left( v_0 s_i + \mu f_{\phi}(z_i) \right) \rho_{\text{tot}}([\phi, \Pi], z, s, t) \right) + \alpha \left( \rho_{\text{tot}}([\phi, \Pi], z, - s_i,t) -  \rho_{\text{tot}}([\phi, \Pi], z, s_i,t)\right) \right] \nonumber \\
    & \qquad - m^{-1} \oint \id r' \ \Pi(r',t) \  \frac{\delta \rho_{\text{tot}}}{\delta \phi(r',t)}([\phi, \Pi], z,s,t) \nonumber \\
    & \qquad -  \oint \id r' \ \left(m c^2\frac{\partial^2 \phi}{\partial r'^2}(r',t) -\kappa_0 \phi(r',t) -\zeta_\phi\sum_i^N G(r' - z_i) \right) \frac{\delta \rho_{\text{tot}}}{\delta \Pi(r',t)}  ([\phi, \Pi], z,s,t) \nonumber
\end{align}
with $\oint = \int_0^L$, defining the forward generator $\mathcal{L}^\dagger$. It can be rewritten in the time scale separation form with the small quantities $\ve_\phi, \ve_\Pi$ that appeared in Eq. (6) of the main text
 \begin{align}\label{schematic L dagger}
   \mathcal{L}^\dagger & = \mathcal{L}_{z, s}^{\dagger} + \ve_\phi \ \mathcal{L}_\phi^{\dagger} + \ve_{\Pi}  \mathcal{L}_{\Pi}^{\dagger}, \qquad \ve_\phi = \frac{\Pi_c L}{m v_0 \phi_0}, \qquad \ve_{\Pi} = \frac{m c^2 \phi_0}{L v_0 \Pi_c} = \frac{Y \phi_0}{L v_0 \Pi_c}
    \end{align}
where
    \begin{align}
    &\mathcal{L}_{z, s}^{\dagger} \rho_{\text{tot}}([\phi, \Pi],z, s, t) = \sum_{i}^N \mathcal{L}_{z_i, s_i}^{\dagger} \rho_{\text{tot}}([\phi, \Pi],z, s, t)  \nonumber \\
    & =    \sum_{i}^N \Big[ -\frac{\partial}{\partial z_i} \left( \left( v_0 s_i + \mu f_{\phi}(z_i) \right) \rho_{\text{tot}}([\phi, \Pi], z, s, t) \right) + \alpha \left( \rho_{\text{tot}}([\phi, \Pi], z, - s_i,t) -  \rho_{\text{tot}}([\phi, \Pi], z, s_i,t)\right) \Big] \nonumber \\
    & \mathcal{L}_\phi^{\dagger} \rho_{\text{tot}}([\phi, \Pi],z, s, t) = - v_0 \oint \frac{\id r'}{L} \ \frac{\Pi(r',t)}{\Pi_c} \frac{\delta \rho_{\text{tot}}}{\delta \left(  \phi(r',t)/\phi_0 \right)} ([\phi, \Pi], z,s,t) \label{Lphi dagger} \\
    &  \mathcal{L}_\Pi^{\dagger} \rho_{\text{tot}}([\phi, \Pi],z, s, t) = - v_0 \oint \frac{\id r'}{L} \ \left(\frac{\partial^2 (\phi(r',t)/\phi_0)}{\partial (r'/L)^2} - \frac{\kappa_0 L^2}{mv^2} \frac{\phi(r',t)}{\phi_0} -\frac{ \zeta_\phi \ L}{m c^2 \phi_0} \sum_i^N L \ G(r' - z_i) \right)  \label{LPidagger} \\
    & \hspace{10 cm} \cdot \frac{\delta \rho_{\text{tot}}}{\delta (\Pi(r',t)/\Pi_c)}([\phi, \Pi], z,s,t)  \nonumber
\end{align}
This splitting of the generator invites a time-scale separation of the dynamics since $\ve_\phi, \ve_\Pi \ll 1$ while for $v_0 $ sufficiently large, $\mathcal{L}_{z,s}^{\dagger}/\mathcal{L}_\phi^\dagger, \mathcal{L}_{z,s}^{\dagger}/\mathcal{L}_\Pi^\dagger \sim O(1)$. 
The forward generator has a similar form $ \mathcal{L} = \mathcal{L}_{z, s} +\ve_\phi  \mathcal{L}_\phi + \ve_\Pi \mathcal{L}_\Pi$ where the particle generator $\mathcal{L}_{z, s}^{\dagger}$ contains the nonconservative effects from the underlying environment and active driving while the $\phi, \Pi$ dynamics is conservative such that $\mathcal{L}_\phi = -  \mathcal{L}_\phi^{\dagger}, \mathcal{L}_\Pi = - \mathcal{L}_\Pi^{\dagger}$. 

\section{Derivation of the induced Fokker-Planck equation for \texorpdfstring{$\phi$}{ϕ}}\label{appendix fokker plank phi}
The generator \eqref{schematic L dagger} has the same form as in \cite{pei2024inducedfrictionprobemoving}, due to the assumed timescale separation of the field and the particles, characterized by the small constants $\ve$. We must track terms to order $\ve^2$ and focus on the behavior at the field time scale $t \sim O(\ve^{-2})$. The resulting dynamics for the field is a Markov diffusive process, which can be expressed by a formal Fokker-Planck equation for the reduced distribution 
\begin{align}
    \tilde{\rho}([\phi, \Pi],t) &= \sum_{\vec{s}} \oint \id \vec{z} \ \rho_{\text{tot}}([\phi, \Pi], z, s)=    \sum_{s_1 = \pm 1} ... \sum_{s_N = \pm 1} \oint \id z_1 ... \id z_N  \ \rho_{\text{tot}}([\phi, \Pi], z, s) \nonumber
\end{align}
To derive the Fokker-Planck equation for $ \tilde{\rho}$, one starts from the Nakajima-Zwanzig equation \cite{nakajima1958quantum, zwanzig, zwanzig2001nonequilibrium, pei2024inducedfrictionprobemoving} for the total distribution $\rho_{ \text{tot}}$,
\begin{align}\label{projected fokker planck}
  & \frac{\partial}{\partial t} \mathcal{P}^{\dagger} \rho_{ \text{tot}}(t) \nonumber  \\
   &=  \mathcal{P}^{\dagger} \mathcal{L}^{\dagger} e^{t \mathcal{Q}^{\dagger} \mathcal{L}^{\dagger}} \mathcal{Q}^{\dagger} \rho_{ \text{tot}}(0) + \mathcal{P}^{\dagger} \mathcal{L}^{\dagger}  \mathcal{P}^{\dagger} \rho_{ \text{tot}}(t)  + \int_0^t \id \tau \ \mathcal{P}^{\dagger} \mathcal{L}^{\dagger} e^{\tau \mathcal{Q}^{\dagger} \mathcal{L}^{\dagger}} \mathcal{Q}^{\dagger} \mathcal{L}^{\dagger} \mathcal{P}^{\dagger}  \rho_{ \text{tot}}(t-\tau) 
\end{align}
with initial condition $\rho_{ \text{tot}}(0)$. Here, $\mathcal{P}^{\dagger}$ represents the projector operator
\begin{equation*}
    \mathcal{P}^{\dagger} h([\phi, \Pi], z, s) = \rho_\phi(z, s) \sum_{\zvec{s}'} \oint   \id \zvec{z}' \ h([\phi, \Pi], z', s')
\end{equation*}
which traces out the medium $  \text{Tr}_{z, s} = \sum_{\zvec{s}'} \oint \id \zvec{z}' $ and replaces the medium distribution with the pinned (or Born-Oppenheimer) distribution $\rho_{\phi}(z, s)$ satisfying 
\begin{align}\label{def rho_phi}
 0 =  \mathcal{L}_{z, s}^{\dagger} \rho_\phi(z, s) =    \sum_{i}^N \Big[ - \frac{\partial}{\partial z_i} \left( \left( v_0 s_i + \mu f_{\phi(t)}(z_i) \right) \rho_{\phi}(z, s) \right) + \alpha \left( \rho_{\phi}(z, - s_i) -  \rho_{\phi}(z, s_i)\right) \Big] 
\end{align}
where $\rho_{\phi}(z, - s_i)$ flips only the $i$th spin $s_i$. 
The conjugate projection operator $\mathcal{P}$ is defined on functions $g = g([\phi, \Pi], z, s)$ by
\begin{align*}
    (\mathcal{P} \ g)[\phi, \Pi] = \sum_{\vec{s}} \oint \id \vec{z} \ g([\phi, \Pi], z, s) \ \rho_\phi(z, s) = \langle g \rangle_{\phi}^{\text{BO}} 
\end{align*}
and motivates the Born-Oppenheimer average $\langle \cdot \rangle_{\phi}^{\text{BO}}$ in Eq. (9) in the main text. Furthermore, from the definitions of $\mathcal{P}, \mathcal{P}^{\dagger}$ and \eqref{def rho_phi}, it follows that
\begin{align}
    \mathcal{L}_{z,s} \mathcal{P}& = 0, \qquad  \mathcal{L}_{z,s}^{\dagger}  \mathcal{P}^{\dagger}  = 0, \qquad \text{Tr}_{z, s} \mathcal{P}^{\dagger} = \text{Tr}_{z, s} \label{P L = 0} \\
   \mathcal{P}^{\dagger} \mathcal{L}_{z,s}^{\dagger} &= 0, \qquad  \mathcal{P} \ \mathcal{L}_{z,s}  = 0 \label{P L = 0 2}
\end{align}
In what follows, we use the orthogonal projection operator $\mathcal{Q} = 1 - \mathcal{P}$ to $\mathcal{P}$ with conjugate $\mathcal{Q}^{\dagger} = 1 - \mathcal{P}^{\dagger}$. \\

We wish to show
\begin{align}\label{reduced master eq phi}
 \frac{\partial \tilde{\rho}}{\partial t}([\phi, \Pi)],t) &=- \oint \id r'  \left( \frac{\delta}{\delta \phi(r',t)} \left[\tilde{\rho}([\phi, \Pi],t) \tilde{K}_\phi \right] + \frac{\delta}{\delta \Pi(r',t)} \left[ \tilde{\rho}([\phi, \Pi],t) \tilde{K}_\Pi \right] \right) \\
    & \qquad - \oint \id r'  \frac{\delta}{\delta \Pi(r',t)} \left[\tilde{\rho}([\phi, \Pi],t) \oint \id u' \left(  -\tilde{\nu}(r', u', [\phi]) \frac{\partial \phi}{\partial t}+  \frac{\delta \tilde{B}}{\delta \Pi(u',t)} \right)  \right] \label{reduced master eq phi line 2} \\
    & \qquad +\oint\id r' \ \id u' \frac{\delta}{\delta \Pi(r')} \frac{\delta}{\delta \Pi(u')}  \left[ \tilde{\rho}([\phi, \Pi],t) \ \tilde{B}(r',u', [\phi]) \right] \label{reduced master eq phi line 3}
\end{align}

\subsection{Expansion of the Nakajima-Zwanzig equation in \texorpdfstring{$\ve$}{ε}}
Under the time scale separation, we only track the terms up to $O(\ve^2)$ in \eqref{projected fokker planck}, \textit{i.e.} we concentrate on the long-time behavior $t \in [\frac{T}{\ve^2}, \infty)$ with $\ve \to 0$. That is also the time scale of the friction and noise effects. Analyzing each term separately:
\begin{itemize}
\item The first term on the right-hand side denotes the effect of the initial distribution $ \rho_{ \text{tot}}(0)$, which vanishes if the initial distribution can be decomposed as the system distribution times the pinned distribution, \textit{i.e.}
\begin{align*}
    &\rho_{ \text{tot}}(0) = \tilde{\rho}([\phi, \Pi],0) \cdot \rho_{ \phi}(z, s, 0) \Longrightarrow \mathcal{Q}^{\dagger} (\tilde{\rho}([\phi, \Pi],0) \cdot \rho_{ \phi}(z, s, 0)) =0  
\end{align*}
More generally, since we focus on the long-time behavior, this term can be neglected up to $O(\ve^2$) as well since $\mathcal{L}$ has negative eigenvalues leading to an exponential relaxation, \cite{Kampen1992}.
\item Using the projector operator identity \eqref{P L = 0 2}, the second term is of order $O(\ve^1)$ and equal to
\begin{equation*}
  \ve_\phi \mathcal{P}^{\dagger} \mathcal{L}_{\phi}^{\dagger}  \mathcal{P}^{\dagger} \rho_{ \text{tot}}(t) + \ve_\Pi \mathcal{P}^{\dagger} \mathcal{L}_{\Pi}^{\dagger}  \mathcal{P}^{\dagger} \rho_{ \text{tot}}(t)
\end{equation*}
\item Finally, using \eqref{P L = 0} together with the estimates
\begin{align*}
   \mathcal{P}^{\dagger} \rho_{ \text{tot}}(t-\tau) = \mathcal{P}^{\dagger} (1+ O(\ve)) \rho_{ \text{tot}}(t), \qquad 
     e^{\tau \mathcal{Q}^{\dagger} \mathcal{L}^{\dagger}} = 
     \left(e^{\tau  \mathcal{L}_{z,s}^\dagger} - \mathcal{P}^{\dagger} \right) (1 + O(\ve))
\end{align*}
the last term is of order $O(\ve^2)$ (and higher)
\begin{align*}
    &\int_0^t \id \tau \Bigg[ \ve_\phi^2 \ \mathcal{P}^{\dagger} \mathcal{L}_{\phi}^{\dagger} \left(e^{\tau  \mathcal{L}_{z,s}^\dagger} - \mathcal{P}^{\dagger} \right) \mathcal{L}_{\phi}^{\dagger} \mathcal{P}^{\dagger}   \rho_{ \text{tot}}(t)  + \ve_\Pi^2 \ \mathcal{P}^{\dagger} \mathcal{L}_{\Pi}^{\dagger} \left(e^{\tau  \mathcal{L}_{z,s}^\dagger} - \mathcal{P}^{\dagger} \right) \mathcal{L}_{\Pi}^{\dagger} \mathcal{P}^{\dagger}   \rho_{ \text{tot}}(t) \\
    & + \ve_\phi \ve_\Pi \ \mathcal{P}^{\dagger} \mathcal{L}_{\phi}^{\dagger} \left(e^{\tau  \mathcal{L}_{z,s}^\dagger} - \mathcal{P}^{\dagger} \right) \mathcal{L}_{\Pi}^{\dagger} \mathcal{P}^{\dagger}   \rho_{ \text{tot}}(t)  + \ve_\phi \ve_\Pi \ \mathcal{P}^{\dagger} \mathcal{L}_{\Pi}^{\dagger} \left(e^{\tau  \mathcal{L}_{z,s}^\dagger} - \mathcal{P}^{\dagger} \right) \mathcal{L}_{\phi}^{\dagger} \mathcal{P}^{\dagger}   \rho_{ \text{tot}}(t) + O(\ve^3) \Bigg]
\end{align*}
Concentrating on the long-time behavior for $t \geq \frac{T}{\ve^2}$, the Markov approximation can be made. This amounts to setting the upper limit of the time integral in \eqref{projected fokker planck} to infinity $\int_0^t \to \int_0^\infty$. As shown below, the integrands can be expressed as time-dependent correlation functions in the medium only, which do not depend on $\ve$. Therefore, this Markov approximation for the field dynamics will be valid for $t \geq \frac{T}{\ve^2}$. 
\end{itemize}
Equation \eqref{projected fokker planck} is still for the total distribution $\rho_{ \text{tot}}$. We now take the integral over the $z$ particles on both sides to obtain the equation for the reduced distribution $\tilde{\rho} = \text{Tr}_{z, s} \left( \rho_{ \text{tot}} \right)$
\begin{align}
    \frac{\partial}{\partial t} \tilde{\rho}(t) &= \ve_\phi \ \text{Tr}_{z, s}\left( \mathcal{L}_{\phi}^{\dagger}  \mathcal{P}^{\dagger} \rho_{ \text{tot}}(t) \right) +  \ve_\Pi \ \text{Tr}_{z, s}\left( \mathcal{L}_{\Pi}^{\dagger}  \mathcal{P}^{\dagger} \rho_{ \text{tot}}(t) \right) \label{first order terms appendix} \\
     & \qquad + \ve_\phi^2 \int_0^{\infty} \id \tau \ \text{Tr}_{z, s} \left( \mathcal{L}_\phi^{\dagger} \left(e^{\tau  \mathcal{L}_{z,s}^\dagger} - \mathcal{P}^{\dagger} \right) \mathcal{L}_\phi^{\dagger} \mathcal{P}^{\dagger}  \rho_{ \text{tot}}(t) \right) \label{ve_phi^2}\\
      & \qquad + \ve_{\phi} \ve_\Pi \int_0^{\infty} \id \tau \  \text{Tr}_{z, s} \left( \mathcal{L}_{\phi}^{\dagger} \left(e^{\tau  \mathcal{L}_{z,s}^\dagger} - \mathcal{P}^{\dagger} \right) \mathcal{L}_{\Pi}^{\dagger} \mathcal{P}^{\dagger}   \rho_{ \text{tot}}(t) \right) \label{ve_phi ve_Pi}\\
     & \qquad + \ve_\Pi^2 \int_0^{\infty} \id \tau \ \text{Tr}_{z, s} \left( \mathcal{L}_\Pi^{\dagger} \left(e^{\tau  \mathcal{L}_{z,s}^\dagger} - \mathcal{P}^{\dagger} \right) \mathcal{L}_\Pi^{\dagger} \mathcal{P}^{\dagger}  \rho_{ \text{tot}}(t) \right) \label{ve_Pi^2} \\
     & \qquad+ \ve_\Pi \ve_{\phi} \int_0^{\infty} \id \tau \  \text{Tr}_{z, s} \left( \mathcal{L}_{\Pi}^{\dagger} \left(e^{\tau  \mathcal{L}_{z,s}^\dagger} - \mathcal{P}^{\dagger} \right) \mathcal{L}_{\phi}^{\dagger} \mathcal{P}^{\dagger} \rho_{ \text{tot}}(t) \right) \label{ve_Pi ve_phi}
\end{align}
Next, we analyze the right-hand side order by order.

\subsection{Leading order \texorpdfstring{$O(\ve^1)$}{O(ε¹)}}
Starting with the first-order terms \eqref{first order terms appendix}, we apply the projector $\mathcal{P}^{\dagger}$ on $\mathcal{L}_\phi^{\dagger}, \mathcal{L}_\Pi^\dagger$ in \eqref{Lphi dagger}, \eqref{LPidagger}, yielding
\begin{align}
    \ve_\phi \text{Tr}_{z, s}\left( \mathcal{L}_{\phi}^{\dagger}  \mathcal{P}^{\dagger} \rho_{ \text{tot}}(t) \right) 
    & = -  \ve_\phi \ v_0 \phi_0 \sum_{\vec{s}} \oint \id \vec{z} \ \id r' \ \frac{\Pi(r',t)}{\Pi_c} \frac{\delta}{\delta \phi(r',t)} \left[\rho_\phi(z, s) \ \tilde{\rho}([\phi, \Pi],t) \right] \nonumber \\
    & = - \ \oint \id r' \  \frac{\delta}{\delta \phi(r',t)} \left[\tilde{\rho}([\phi, \Pi],t) \ \left \langle \mathcal{L}\phi(r',t) \right \rangle_\phi^{\text{BO}}  \right] \label{Kphi term appendix} 
\end{align}
where one recognizes the first term in \eqref{reduced master eq phi} with induced force
\begin{equation*}
     \tilde{K}_\phi(r,t) = \left \langle \mathcal{L} \phi(r,t) \right \rangle_\phi^{\text{BO}} = \left \langle \ve_\phi \mathcal{L}_\phi \phi(r,t) \right \rangle_\phi^{\text{BO}} = m^{-1} \Pi(r,t)
\end{equation*}
Similarly, for the second term, 
\begin{align}
  \ve_\Pi \text{Tr}_{z, s}\left( \mathcal{L}_{\Pi}^{\dagger}  \mathcal{P}^{\dagger} \rho_{ \text{tot}}(t) \right)  & = - \oint \id r' \  \frac{\delta}{\delta \Pi(r',t)} \left[\tilde{\rho}([\phi, \Pi],t) \left \langle  \mathcal{L} \Pi \right \rangle_\phi^{\text{BO}}(\phi(r',t), r')) \right]  \label{KPi term appendix}
\end{align}
This result corresponds to the second term in \eqref{reduced master eq phi} with the streaming term
\begin{align}
    \tilde{K}_\Pi(r,t) &= \left \langle  \mathcal{L} \Pi \right \rangle_\phi^{\text{BO}} = \ve_\Pi \left \langle  \mathcal{L}_\Pi \Pi \right \rangle_\phi^{\text{BO}}  =   m c^2\frac{\partial^2 \phi}{\partial r^2}(r,t) -\kappa_0 \phi(r,t) 
    - \zeta_\phi \sum_{i = 1}^N \left \langle G(r - z_i(t)) \right \rangle_\phi^{\text{BO}}  \label{streaming term appendix}
\end{align}
These are the only terms that appear at order $O(\ve^1)$. 

\subsection{Next-to-leading order \texorpdfstring{$O(\ve^2)$}{O(ε²)}}
At next-to-leading order, we compute the four terms \eqref{ve_phi^2}--\eqref{ve_phi ve_Pi}--\eqref{ve_Pi^2}--\eqref{ve_Pi ve_phi}. 
\subsubsection{\texorpdfstring{$\ve_\phi^2$}{εₚₕᵢ²} and \texorpdfstring{$\ve_\phi  \ve_\Pi$}{εₚₕᵢ εₚᵢ} contributions}
Following \eqref{Kphi term appendix}, the $\ve_\phi^2$ term \eqref{ve_phi^2} becomes
\begin{align*}
    &\ve_\phi^2 \int_0^{\infty} \id \tau \ \text{Tr}_{z, s} \left( \mathcal{L}_\phi^{\dagger} \left(e^{\tau  \mathcal{L}_{z,s}^\dagger} - \mathcal{P}^{\dagger} \right) \mathcal{L}_\phi^{\dagger} \mathcal{P}^{\dagger}  \rho_{ \text{tot}}(t) \right) \\
    & = - \ve_\phi \int_0^{\infty} \id \tau \sum_{\vec{s}} \oint \id \vec{z} \ \id r' \ \mathcal{L}_\phi^{\dagger} \left(e^{\tau  \mathcal{L}_{z,s}^\dagger} - \mathcal{P}^{\dagger} \right) \ \mathcal{L} \phi(r',t) \ \ \frac{\delta}{\delta \phi(r',t)} \left[ \rho_\phi(z, s) \ \tilde{\rho}([\phi, \Pi],t) \right]
\end{align*}
To deal with the operator $\mathcal{L}_\phi^{\dagger} \left(e^{\tau  \mathcal{L}_{z,s}^\dagger} - \mathcal{P}^{\dagger} \right)$, we introduce a delta-functional $\delta(\phi- \varphi)$, a functional integral over $\varphi$ and move the operator to act on the delta-function with its conjugate $\left(e^{\tau  \mathcal{L}_{z,s}} - \mathcal{P} \right) \mathcal{L}_\phi $,
\begin{align*}
    & = - \ve_\phi \int_0^{\infty} \id \tau \sum_{\vec{s}} \oint \id \vec{z} \ \id r' \ \id [\varphi] \ \Big( \left(e^{\tau  \mathcal{L}_{z,s}} - \mathcal{P} \right) \mathcal{L}_{\varphi} \ \delta(\phi - \varphi) \Big) \ \mathcal{L} \varphi(r',t) \ \frac{\delta}{\delta \varphi(r',t)} \left[  \rho_{\varphi}(z, s) \ \tilde{\rho}([\varphi, \Pi],t) \right]
\end{align*}
Using $\mathcal{L}_\varphi = - \mathcal{L}_\varphi^{\dagger}$ and \eqref{Lphi dagger}, the middle term $\left(e^{\tau  \mathcal{L}_{z,s}} - \mathcal{P}' \right) \mathcal{L}_{\varphi} \ \delta(\phi - \varphi)$ becomes
\begin{align}
     \ve_\phi \left(e^{\tau  \mathcal{L}_{z,s}} - \mathcal{P} \right) \mathcal{L}_{\varphi} \ \delta(\phi - \varphi) &= \ve_\phi \ \frac{v_0 \phi_0}{\Pi_c} \oint \id u' \left(e^{\tau  \mathcal{L}_{z,s}} - \mathcal{P} \right) \ \Pi(u',t) \  \frac{\delta}{\delta \varphi(u',t)} \left[ \delta(\phi - \varphi) \right] \nonumber \\
    & = \ve_\phi \ \frac{v_0 \phi_0}{\Pi_c} \oint \id u'  \left( \Pi(u',t)-  \Pi(u',t) \right)\  \frac{\delta}{\delta \varphi(u',t)} \left[ \delta(\phi - \varphi) \right] \nonumber  \\
    & = 0 \nonumber
\end{align}
Hence, \eqref{ve_phi^2} does not contribute. A similar calculation shows that the $\ve_\phi \ve_\Pi$ term \eqref{ve_phi ve_Pi} vanishes as well. Indeed, these terms were absent in \eqref{reduced master eq phi line 2}. 

\subsubsection{\texorpdfstring{$\ve_\Pi^2$}{εₚᵢ²} contribution}
Following the result \eqref{KPi term appendix}, one finds for \eqref{ve_Pi^2}
\begin{align*}
    &\ve_\Pi^2 \int_0^{\infty} \id \tau \ \text{Tr}_{z, s} \left( \mathcal{L}_\Pi^{\dagger} \left(e^{\tau  \mathcal{L}_{z,s}^\dagger} - \mathcal{P}^{\dagger} \right) \mathcal{L}_\Pi^{\dagger} \mathcal{P}^{\dagger}  \rho_{ \text{tot}}(t) \right) \\
    & = - \ve_\Pi \int_0^{\infty} \id \tau \sum_{\vec{s}} \oint \id \vec{z} \ \id r' \ \mathcal{L}_\Pi^{\dagger} \left(e^{\tau  \mathcal{L}_{z,s}^\dagger} - \mathcal{P}^{\dagger} \right) \ \mathcal{L} \Pi(\phi(r',t), r', z(t)) \ \rho_\phi(z, s) \ \frac{\delta \tilde{\rho}}{\delta \Pi(r',t)} ([\phi, \Pi],t) \\
     &  = - \ve_\Pi \int_0^{\infty} \id \tau \sum_{\vec{s}} \oint \id \vec{z} \ \id r' \ \id [\varpi] \ \Big( \left(e^{\tau  \mathcal{L}_{z,s}} - \mathcal{P} \right) \mathcal{L}_{\varpi} \ \delta(\Pi - \varpi) \Big) \ \mathcal{L} \varpi(r', t , z(t)) \ \rho_{\phi}(z, s) \ \frac{\delta \tilde{\rho}}{\delta \varpi(r',t)}([\phi, \varpi],t) 
\end{align*}
where we introduced the delta-functional again. For notational simplicity, we also suppressed the dependence on $\phi$ in $\mathcal{L} \varpi(r', t, z(t)) := \mathcal{L} \varpi(\phi(r',t), r', z(t)) $.
Using $\mathcal{L}_\Pi = - \mathcal{L}_\Pi^{\dagger}$ combined with \eqref{LPidagger}, the term acting on the delta-functional becomes
\begin{align}
   &  \ve_\Pi \left(e^{\tau  \mathcal{L}_{z,s}} - \mathcal{P}' \right) \mathcal{L}_{\varpi} \ \delta(\Pi - \varpi)
     = \oint \id u'  \ \left(e^{\tau  \mathcal{L}_{z,s}} - \mathcal{P} \right) \ \mathcal{L} \varpi(u', t , z(t)) \  \frac{\delta}{\delta \varpi(u',t)} \left[ \delta(\Pi - \varpi) \right] \nonumber \\
    &  = \oint \id u' \  \left(\mathcal{L} \varpi(u', t , z(t+\tau)) - \left \langle \mathcal{L} \varpi \right \rangle_\phi^{\text{BO}}(u',t) \right)  \frac{\delta}{\delta \varpi(u',t)} \left[ \delta(\Pi - \varpi) \right] \nonumber
\end{align}
In the last line, we have replaced $e^{\tau   \mathcal{L}_{z,s}} \mathcal{L} \varpi(r, t, z(t)) $ with the time-evolution value in the pinned dynamics, $\mathcal{L} \varpi(r, t, z(t+\tau))$. Now that all operators have been applied, we eliminate the delta-functional through integration by parts,
\begin{align}
    & \ve_\Pi^2 \int_0^{\infty} \id \tau \ \text{Tr}_{z, s} \left( \mathcal{L}_\Pi^{\dagger} \left(e^{\tau  \mathcal{L}_{z,s}^\dagger} - \mathcal{P}^{\dagger} \right) \mathcal{L}_\Pi^{\dagger} \mathcal{P}^{\dagger}  \rho_{ \text{tot}}(t) \right) \label{ve_Pi^2 intermediate} \\
    & = \oint \id u' \ \id r' \frac{\delta}{\delta \Pi(u',t)}  \left[ \frac{\delta  \tilde{\rho}}{\delta \Pi(r',t)} ([\phi, \Pi],t)\int_0^{\infty} \id \tau \ \left \langle \delta L \Pi(u',t,z(t+ \tau)) \   \ \mathcal{L} \Pi(r', t, z(t)) \right \rangle_\phi^{\text{BO}} \right] \nonumber \\
    &\delta L \Pi(u',t,z(t+ \tau))  =  \mathcal{L} \Pi(u', t, z(t+\tau)) - \left \langle \mathcal{L} \Pi \right \rangle_\phi^{\text{BO}}(u',t) \label{delta L Pi}
\end{align}
The expectation $\left \langle \delta L \Pi(u',t,z(t+ \tau)) \   \ \mathcal{L} \Pi(r', t, z(t)) \right \rangle_\phi^{\text{BO}}$ can be rewritten following the identities
\begin{align}
     \left \langle \mathcal{L} \Pi(u', t, z(t+\tau))  \right \rangle_\phi^{\text{BO}} & = \left \langle \mathcal{L} \Pi  \right \rangle_\phi^{\text{BO}}(u', t), \qquad  \left \langle X \ ; \ Y  \right \rangle_\phi^{\text{BO}} =  
      \left \langle (X - \left \langle X \right \rangle_\phi^{\text{BO}}) \cdot Y  \right \rangle_\phi^{\text{BO}} \label{covariance result}
\end{align}
where we use \eqref{P L = 0} in the first equation.  Therefore,
\begin{align*}
&\ve_\Pi^2 \int_0^{\infty} \id \tau \ \text{Tr}_{z, s} \left( \mathcal{L}_\Pi^{\dagger} \left(e^{\tau  \mathcal{L}_{z,s}^\dagger} - \mathcal{P}^{\dagger} \right) \mathcal{L}_\Pi^{\dagger} \mathcal{P}^{\dagger}  \rho_{ \text{tot}}(t) \right)  \\
   & = \oint \id u' \ \id r' \frac{\delta}{\delta \Pi(u',t)}  \left[ \frac{\delta  \tilde{\rho}}{\delta \Pi(r',t)} ([\phi, \Pi],t) \int_0^{\infty} \id \tau \ \left \langle \mathcal{L} \Pi(s) \ ; \ \mathcal{L} \Pi \right \rangle_\phi^{\text{BO}} \right] \nonumber \\
    & = \oint \id u' \ \id r' \frac{\delta}{\delta \Pi(u',t)} \frac{\delta}{\delta \Pi(r',t)}  \left[ \tilde{\rho}([\phi, \Pi],t) \int_0^{\infty} \id \tau \ \Big \langle \mathcal{L} \Pi(u', t, z(t+\tau)) \ ; \ \mathcal{L} \Pi(r',t,z(t)) \Big \rangle_\phi^{\text{BO}} \right] \\
    & - \oint \id u' \  \frac{\delta}{\delta \Pi(u',t)}  \left[ \tilde{\rho}([\phi, \Pi],t) \int \id r' \frac{\delta}{\delta \Pi(r',t)} \left(  \int_0^{\infty} \id \tau \ \Big \langle \mathcal{L} \Pi(u', t, z(t+\tau)) \ ; \ \mathcal{L} \Pi(r',t,z(t)) \Big\rangle_\phi^{\text{BO}} \right) \right] \nonumber
\end{align*}
which is the final result for \eqref{ve_Pi^2}. One recognizes the two contributions of the noise coefficient $\tilde{B}_{\Pi}$ in \eqref{reduced master eq phi line 2}-\eqref{reduced master eq phi line 3} where
\begin{align}
    \tilde{B}(r,u, [\phi])   &= \int_0^{\infty} \id \tau \ \Big \langle \mathcal{L} \Pi(r, t, z(t+\tau)) \ ; \ \mathcal{L} \Pi(u,t,z(t)) \Big\rangle_\phi^{\text{BO}} \nonumber \\
    & =\zeta_\phi^2 \sum_i^N \sum_j^N \int_0^{\infty} \id \tau \ \Big \langle  G(r - z_i(t+\tau)) \ ; \  G(u - z_j(t)) \Big \rangle_{\phi}^{\text{BO}} \label{reduced form D 2}
\end{align}
\subsubsection{\texorpdfstring{$\ve_\Pi \ \ve_\phi$}{εₚᵢ εₚₕᵢ} contribution}
Following similar steps that lead to \eqref{ve_Pi^2 intermediate}, the term \eqref{ve_Pi ve_phi} becomes
\begin{align*}
     &\ve_{\Pi} \ve_\phi \int_0^{\infty} \id \tau \ \text{Tr}_{z, s} \left(\mathcal{L}_{\Pi}^{\dagger} \left(e^{\tau  \mathcal{L}_{z,s}^\dagger} - \mathcal{P}^{\dagger} \right) \mathcal{L}_{\phi}^{\dagger} \mathcal{P}^{\dagger}   \rho_{ \text{tot}}(t) \right) \\
     &= \oint \id u' \ \frac{\delta}{\delta \Pi(u',t)} \Bigg[ \int_0^{\infty} \id \tau \sum_{\vec{s}} \oint \id \vec{z} \ \id r' \ \delta L \Pi(u',t,z(t+\tau)) \  \mathcal{L} \phi(r',t) \ \frac{\delta}{\delta \phi(r',t)} \left[\rho_{\phi}(z, s) \ \tilde{\rho}([\phi, \Pi],t) \right]  \Bigg]
\end{align*}
Expanding the last functional derivative yields two terms $T_1, T_2$, where $T_1$ equals
\begin{align*}
   & T_1  = \oint \id u' \ \id r' \frac{\delta}{\delta \Pi(u',t)}  \Bigg[ \frac{\delta}{\delta \phi(r',t)} \left( \tilde{\rho}([\phi, \Pi],t) \right)  \ \mathcal{L} \phi(r',t) \int_0^{\infty} \id \tau \sum_{\vec{s}} \oint \id \vec{z} \ \rho_{\phi}(z, s) \ \delta L \Pi(u',t,z(t+\tau)) \Bigg] \nonumber \\
    & = \oint \id u' \ \id r' \frac{\delta}{\delta \Pi(u',t)}  \left[ \frac{\delta}{\delta \phi(r',t)} \left( \tilde{\rho}([\phi, \Pi],t) \right) \ \mathcal{L} \phi(r',t) \int_0^{\infty} \id \tau \  \left \langle \delta L \Pi(u',t,z(t+\tau))  \right \rangle_\phi^{\text{BO}}   \right] \nonumber \\
    & = 0
\end{align*}
In the last line, we have used $\left \langle \delta L \Pi(u',t,z(t+\tau))  \right \rangle_\phi^{\text{BO}}$ following \eqref{covariance result}. \\
Next, for $T_2$,
\begin{align*}
    T_2 &=  \oint \id u' \ \id r' \frac{\delta}{\delta \Pi(u',t)}  \left[  \tilde{\rho}([\phi, \Pi],t) \int_0^{\infty} \id \tau \   \delta L \Pi(u',t,z(t+\tau))  \ \mathcal{L} \phi(r',t) \frac{\delta \rho_\phi}{\delta \phi(r',t)} (z, s) \right] \nonumber \\
    &=  \oint \id u' \ \id r' \frac{\delta}{\delta \Pi(u',t)}  \Bigg[  \tilde{\rho}([\phi, \Pi],t) \  \int_0^{\infty} \id \tau \ \Big \langle  \delta L \Pi(u',t,z(t+\tau)) \  \mathcal{L}\phi(r',t) \frac{\delta  \log \rho_\phi}{\delta \phi(r',t)}(z, s)   \Big\rangle_\phi^{\text{BO}} \Bigg] \nonumber
\end{align*}
Applying the covariance result \eqref{covariance result} with the definition \eqref{delta L Pi} for $\delta L \Pi$ yields
\begin{align*}
    T_2 & = \oint \id u' \frac{\delta}{\delta \Pi(u',t)}  \Bigg[  \tilde{\rho}([\phi, \Pi],t) \  \int_0^{\infty} \id \tau \oint  \id r' \ \Big \langle \mathcal{L} \Pi(u', t , z(t+\tau)) \ ;  \ \frac{\delta  \log \rho_\phi}{\delta \phi(r',t)}(z, s)   \Big\rangle_\phi^{\text{BO}} \ \mathcal{L}\phi(r',t) \Bigg] 
\end{align*}
which is the final expression for \eqref{ve_Pi ve_phi} and agrees with the friction term $\tilde{\nu}$ in \eqref{reduced master eq phi line 2} with
\begin{align}
     \tilde{\nu}(r, u, [\phi]) &=  \int_0^{\infty} \id \tau  \ \Big \langle \mathcal{L} \Pi(r, t , z(t+\tau)) \ ;  \ \frac{\delta  \log \rho_\phi}{\delta \phi(u,t)}(z, s)   \Big\rangle_\phi^{\text{BO}} \nonumber \\
     & = - \zeta_\phi \sum_i^N \int_0^{\infty} \id \tau \ \left \langle G(r - z_i(t+\tau)) \ ;  \ \frac{\delta \log \rho_\phi}{\delta \phi(u,t)} (z, s)   \right \rangle_\phi^{\text{BO}} \label{reduced form nu s}
\end{align}
The covariance $\langle \cdot \ ; \ \cdot \rangle^{\text{BO}}$ in $\tilde{\nu}$ can be rewritten as a single expectation value
\begin{align}
   \tilde{\nu}(r, u, [\phi]) &=  - \zeta_\phi \sum_i^N \int_0^{\infty} \id \tau \ \left \langle G(r - z_i(t+\tau))  \cdot   \frac{\delta \log \rho_\phi}{\delta \phi(u,t)} (z, s)   \right \rangle_\phi^{\text{BO}} \nonumber \\
    & \qquad + \zeta_\phi  \sum_i^N \int_0^\infty  \id \tau \ \big \langle G(r - z_i(t + \tau))  \big\rangle_\phi^{\text{BO}} \cdot \left \langle \ \frac{\delta \log \rho_\phi}{\delta \phi(u,t)}(z, s)   \right \rangle_\phi^{\text{BO}} \nonumber \\
    & = - \zeta_\phi \sum_i^N \int_0^{\infty} \id \tau \ \left \langle G(r - z_i(t+\tau)) \ \cdot  \ \frac{\delta \log \rho_\phi}{\delta \phi(u,t)} (z, s)   \right \rangle_\phi^{\text{BO}} \label{reduced form nu 2}
\end{align}
where we have used the normalization of $\rho_\phi$ to eliminate the second term since
\begin{align*}
    \left \langle  \frac{\delta \log \rho_\phi}{\delta \phi(u,t)}(z, s)   \right \rangle_\phi^{\text{BO}} & = \sum_{\vec{s}} \int \id \vec{z} \ \frac{\delta \rho_\phi}{\delta \phi(u,t)}(z, s) = \frac{\delta}{\delta \phi(u,t)} \left(1 \right) = 0
\end{align*}
Depending on the context, \eqref{reduced form nu s} or \eqref{reduced form nu 2} will be used. \\

\begin{figure}[ht]
    \centering
    \includegraphics[width=0.8\linewidth]{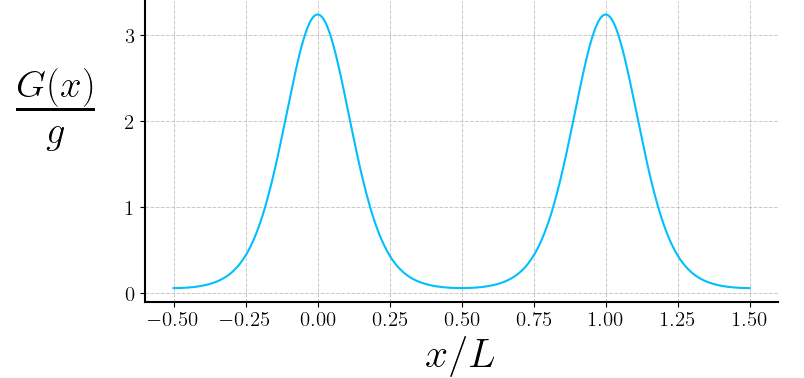}
    \caption{Von Mises distribution for $p = 2$ in Eq. (6) of the main text. The distribution is peaked at $x = 0 = L$.}
    \label{fig von mises}
\end{figure}

Putting all terms together, we obtain the Fokker-Planck equation \eqref{reduced master eq phi}, which is equivalent to the Langevin equation 
\begin{align*}
  \frac{\partial \phi}{\partial t}(r,t) &= \tilde{K}_\phi(\Pi(r,t)) =  m^{-1} \Pi(r,t)\\
\frac{\partial \Pi}{\partial t}(r,t) &= \tilde{K}_\Pi(\phi(r,t), r) - \oint \id u \ \tilde{\nu}(r, u, [\phi]) \ \frac{\partial \phi}{\partial t}(u,t) \\
& + \oint \id u   \ \frac{\delta \tilde{B}}{\delta \Pi(u,t)}(r,u, [\phi]) + \oint \id u \ \tilde{\Gamma}(r,u,[\phi]) \ \xi(u,t)
\end{align*}
with white spacetime noise $\xi(r,t)$, and $\tilde{\Gamma}$ the square root of the noise amplitude $2 \tilde{B}$ as in Eq. (8) and above Eq. (12) in the main text, respectively. As expected, the momentum relation $\partial_t \phi = m^{-1} \Pi$ is unchanged. 
\\Moreover, the derivative $ \frac{\delta B}{\delta \Pi(u,t)}$ vanishes since neither $G$ nor $\rho_\phi$ depends on $\Pi$ (due to our choice of the interaction between the string and particles). 
Consequently, the equations can be combined into the modified Klein-Gordon equation
\begin{align}
   \frac{1}{c^2} \frac{\partial^2 \phi}{\partial t^2}(r,t) &= K_\Pi(\phi(r,t) ,r) - \oint \id u \  \nu(r, u, [\phi]) \ \frac{\partial \phi}{\partial t}(u,t)  + \oint \id u \ \Gamma(r,u,[\phi]) \ \xi(u,t) \label{modified klein gordon}
\end{align}
where $K_\Pi = \tilde{K}_\Pi/Y, \nu = \tilde{\nu}/Y, \gamma= \tilde{\Gamma}/Y$. This result appeared in the main text as Eq. (7) using $k = \zeta_\phi/Y, M^2 = \kappa_0/Y$ and 
\begin{align*}
   K_\Pi(r,t) &= \frac{\partial^2 \phi}{\partial r^2}(r,t) - M^2\phi(r,t) 
    - k \sum_{i = 1}^N \left \langle G(r - z_i(t)) \right \rangle_\phi^{\text{BO}} 
\end{align*}
The function $G(x)$ is periodic $G(x) = G(x + L)$, peaked around $x = 0$, and for $G$ as a function of  $r-z_i(t)$, we need the smallest distance on the circle. Thus we require $G$ to depend on $L$ and to be even around $\frac{L}{2}$ which, for a periodic function, is equivalent to an even function around $0$, \textit{e.g.} $G(x) = h\left(\cos\left(\frac{2 \pi x}{L} \right) \right)$ for some function $h$. In the main text, we consider the von Mises distribution, Eq. (5), plotted in Fig. \ref{fig von mises} 

\section{Stationary distribution run-and-tumble particles at fixed \texorpdfstring{$\phi$}{ϕ}}\label{solution stationary distribution appendix}
The stationary distribution $\rho_\phi$ of the run-and-tumble process at fixed $\phi$ satisfies \eqref{def rho_phi} and factorizes due to the independence of the $z-$particles
\begin{align*}
\rho_\phi(z, s) &= \prod_{i = 1}^N \rho_\phi^i(z_i,s_i), \qquad \rho_\phi^i(z_i,s_i) = \rho_\phi^i(z_i + L,s_i), \qquad \sum_{s_i = \pm 1} \oint \id z_i \ \rho_\phi^i(z_i,s_i) = 1 \\
  0&= -  \ \frac{\partial}{\partial z_i} \left( \left( v_0 s_i + \mu f_{\phi}(z_i) \right) \rho_{\phi}^i(z_i, s_i) \right) + \alpha \left(\rho_{\phi}^i(z_i, -s_i)-  \rho_{\phi}^i(z_i, s_i)\right)
\end{align*}
Its solution is known, \cite{Bena_2003}, which we write in the form
\begin{align}
\rho_\phi^i(z_i, s_i) &= \frac{p_\phi^i(z_i) + s_i q_\phi^i(z_i)}{2}, \qquad  q_\phi^i(z_i) = \frac{J_\phi - \mu f_\phi(z_i) \ p_\phi^i(z_i)}{v_0} \label{stationary rho phi} \\
p_\phi^i(z_i) &= \frac{\psi(z_i)}{v_0^2 - \mu^2 f_{\phi}(z_i)^2}  \Bigg(C_2 - J_\phi \int_{0}^{z_i} \  \frac{(2 \alpha + \mu f'_{\phi}(x))}{\psi(x)} \ \id x \Bigg) \label{stationary p phi} \\
\psi(z_i) &= \exp{\Bigg[ \int_{0}^{z_i} \ I(y) \ \id y \Bigg]} > 0, \qquad I(y)  = 2 \alpha\frac{\mu f_{\phi}(y)}{v_0^2 - \mu^2 f_{\phi}(y)^2} \nonumber 
\end{align}
where the constants $J_\phi, C_2$ follow from the periodic boundary conditions and normalization
\begin{align*}
    J_\phi &= C_2 \ \eta_\phi, \qquad \eta_\phi = \frac{ \left(\psi(L) - 1 \right)}{\oint\left(2 \alpha+ \mu f'_{\phi}(x)\right) \frac{\psi(L)}{\psi(x)} \ \id x} \\
    C_2^{-1} &= \oint \frac{\psi(z)}{v_0^2- \mu^2 f_{\phi}(z)^2} \Bigg( 1 - \zeta \int_{0}^z \frac{\left(2 \alpha+\mu f'_{\phi}(x)\right)}{\psi(x)} \, \  \id x \, \Bigg) \ \id z 
\end{align*}
The current density $J_\phi$ thus vanishes if $\psi(L) = 1$ or
\begin{equation*}
    \oint \frac{\mu f_\phi(x)}{v_0^2 - \mu^2 f_\phi(x)^2} \ \id x = 0
\end{equation*}
\textit{e.g.} if $f_\phi(z)$ is odd around $z = \frac{L}{2}$. Note in particular that $\rho_\phi$ does \textit{not} have the Boltzmann form $e^{- \beta H_\phi}$ as we are out-of-equilibrium.  \\

For small coupling $\zeta_\phi \ll 1$, these expressions simplify to
 \begin{align}
   p_\phi^i(z_i) &= \frac{1}{L} \left[ 1 - zeta_\phi\frac{2 \alpha \mu}{v_0^2} \Bigg( \oint \id u \ G(u-z_i) \ \phi(u) - \frac{1}{L} \oint \id x \ G(x)  \oint \id u \ \phi(u)\Bigg) \right]  + O(\zeta_\phi^2) \label{p phi small coupling} \\
            q_\phi^i(z_i)  &= \zeta_\phi  \frac{\mu}{L v_0} \oint \id u  \ G(u-z_i) \ \frac{\partial \phi}{\partial u}(u)  + O(\zeta_\phi^2) \nonumber\\
            \rho^i_{\phi}(z_i, s_i) & = \frac{1}{2L} \Bigg[ 1 - \zeta_\phi\frac{2 \alpha \mu}{v_0^2} \Bigg( \oint \id u \ G(u-z_i) \ \phi(u) - \frac{1}{L} \oint \id x \ G(x)  \oint \id u \ \phi(u)\Bigg) \nonumber\\
            & \hspace{7 cm}+  \zeta_\phi s_i  \frac{\mu}{v_0} \oint \id u  \ G(u-z_i) \ \frac{\partial \phi}{\partial u}(u) \Bigg]  + O(\zeta_\phi^2) \label{rho phi small coupling}
 \end{align}
 In this limit, the force $f_\phi = \zeta_\phi \tilde{f}_\phi$ is seen as a small perturbation to the active noise $v_0 s_i$, putting us in the realm of linear response theory around nonequilibria.

\section{Streaming term}
In this section, we compute the $O(\ve^1)$ terms in the Fokker-Planck equation, \textit{i.e.}  the streaming term $\left \langle G(r - z_i(t)) \right \rangle_\phi^{\text{BO}}$ in \eqref{streaming term appendix}
\begin{align}
    \left\langle  G(r - z_i)\right\rangle_{\phi}^{\text{BO}} &  =\oint \id z_i \ G(r - z_i) \sum_{s_i = \pm 1} \rho_\phi^i(z_i, s_i) = \oint \id z_i \ G(r - z_i) \ p_\phi^i(z_i) \label{average c}
\end{align}
with $p_\phi^i(z_i) = \rho_\phi^i(z_i, s_i) + \rho_\phi^i(z_i, -s_i)$ the probability density that the probe is at location $z_i$. The full solutions $\rho_\phi$ and $p_\phi$ are given in \eqref{stationary rho phi}--\eqref{stationary p phi} and can be substituted in \eqref{average c}, but the resulting integrals do not reduce to a simple or manageable form. Instead, for weak coupling $\zeta_\phi\ll 1$, we use the form \eqref{p phi small coupling}
\begin{align*}
     \left\langle  G(r - z_i)\right\rangle_{\phi}^{\text{BO}} &= \frac{1}{L} \oint \id z_i \ G(r - z_i)  - \zeta_\phi\frac{2 \alpha \mu}{L v_0^2} \oint \id z_i \ \id u \  G(r-z_i) \ G(u-z_i) \ \phi(u) \\
     & \qquad + \zeta_\phi\frac{2 \alpha \mu}{L^2 v_0^2} \oint \id z_i  \  G(r-z_i) \oint \id x \ G(x)  \oint \id u \ \phi(u)  + O(\zeta_\phi^2)
\end{align*}
To simplify these integrals, we use that for a periodic function $G(x) = G(x + L)$
\begin{align*}
  \forall \  a \in \mathbb{R}: \qquad    \int_a^{a+L} \id x \ G(x) = \oint \id x \ G(x)
\end{align*}
such that
\begin{align*}
    &\oint \id z_i \ G(r-z_i) \ G(u-z_i) 
    = \oint \id x \ G(r-u+x) \ G(x), \qquad 
     \oint \id z_i \ G(r-z_i) 
    = \oint \id x \ G(x)
\end{align*}
Consequently
\begin{align*}
     \left\langle  G(r - z_i)\right\rangle_{\phi}^{\text{BO}} &= \frac{1}{L} \oint \id x \ G(x)  - \zeta_\phi\frac{2 \alpha \mu}{L v_0^2} \oint  \id u \left[ \oint \id x \ \left( G(r-u+x) - \frac{1}{L} \oint \id y \  G(y) \right)\ G(x)  \right] \phi(u) \\
      K_\Pi(r,t)  
     &= \frac{\partial^2 \phi}{\partial r^2}(r,t)- M^2 \phi(r,t) - \zeta_\phi^2 \oint \id u \  \mathcal{M}(r-u) \  \phi(u,t) - \frac{\zeta_\phi}{Y} \frac{N}{L} \oint \id x \ G(x) \\
      \mathcal{M}(u) 
 & = -   \frac{N 2 \alpha \mu}{Y L v_0^2} \oint \id x \ \left( G(u+x) - \frac{1}{L} \oint \id y \  G(y) \right)\ G(x) \nonumber
\end{align*}
as in Eq. (13) of the main text. Going to Fourier space
\begin{align}
    \phi(r,t) &= \sum_{n = - \infty}^{\infty} \phi_n(t) \ e^{i 2 \pi n r/L}, \quad  G(x) = \sum_{n = - \infty}^{\infty} G_n \ e^{i 2 \pi n x/L}, \quad \phi_{-n} = \phi_n^*, \quad G_{-n} = G_n^* \label{Fourier space} \\
       K_\Pi(r,t)  
     &= \sum_{n = - \infty}^{\infty} \left[ \frac{4 \pi^2 n^2}{L^2} \phi_n(t)-  M_{\text{eff},n}^2 \phi_n(t)  - k N \ \delta_{n,0} \ G_n \right] e^{i2 \pi  n r}  \nonumber
\end{align}
with effective mass per mode
\begin{align*}
    M_{\text{eff},n} &= \sqrt{M^2 - \zeta_\phi^2  \frac{N 2 \alpha \mu L}{Yc^2} (1 - \delta_{n,0}) |G_n|^2} + O(\zeta_\phi^3) \\
    & = M \left(1 - (1 - \delta_{n,0}) \zeta_\phi^2  \frac{N 2 \alpha \mu L}{2\kappa_0 v_0^2} |G_n|^2 \right) + O(\zeta_\phi^3)
\end{align*}
in agreement with Eq. (14) in the main text. The conditions on the Fourier coefficients in \eqref{Fourier space} appear since $\phi, G$ are real.

\section{Friction and noise amplitude}
In this section, we compute the $O(\ve^2)$ terms in the Fokker-Planck equation, \textit{i.e.} the friction coefficient $\nu$ \eqref{reduced form nu 2} and noise amplitude $B$ \eqref{reduced form D 2}. 
First, due to the independence of the $z-$ particles, the distribution $\rho_\phi$ factorizes  such that the friction becomes
\begin{align}
     \nu  & = - k\sum_i^N \sum_j^N \int_0^{\infty} \id \tau \ \Big \langle G(r - z_i(t+\tau)) \cdot \frac{\delta \log \rho^j_\phi}{\delta \phi(u,t)} (z_j(t), s_j(t))   \Big \rangle_\phi^{\text{BO}} \label{nu with rho^j}
\end{align}
The independence of the $z-$particles also implies that the covariances in \eqref{reduced form D 2}--\eqref{nu with rho^j} vanish unless $j = i$, \textit{i.e.} the expressions simplify to
\begin{align*}
   \nu  & = - k\sum_i^N \int_0^{\infty} \id \tau \ \Big \langle G(r - z_i(t+\tau)) \cdot  \frac{\delta \log \rho^i_\phi}{\delta \phi(u,t)}(z_i(t), s_i(t))   \Big \rangle_\phi^{\text{BO}} \\
    B &= k^2 \sum_i^N \int_0^{\infty} \id \tau \ \Big \langle  G(r - z_i(t+\tau)) \ ; \  G(u - z_i(t)) \Big \rangle_{\phi}^{\text{BO}}
\end{align*}
Explicitly writing out the covariances yields
\begin{align}
    \nu 
    & =   -k \sum_i^N \int_0^{\infty} \id \tau \oint \id z_i \ \id z_{0,i} \ G(r-z_i) \sum_{s_i, s_{0,i}}  \rho^i_\phi(z_i, s_i,\tau|z_{0,i}, s_{0,i})  \ \frac{\delta \rho_\phi^i}{\delta \phi(u,t)} (z_{0,i}, s_{0,i}) \label{nu explicit} \\
     B 
    & = k^2 \sum_i^N  \int_0^{\infty} \id \tau \Bigg[ \oint \id z_i  \ \id z_{0,i} \ G(r-z_i)  \ G(u - z_{0,i})   \sum_{s_i,s_{0,i}}  \rho_\phi^i(z_i, s_i,\tau|z_{0,i}, s_{0,i}) \  \ \rho^i_\phi(z_{0,i}, s_{0,i})  \nonumber \\
    & \hspace{3 cm} - \left( \oint \id z_i \ G(r-z_i) \sum_{s_i} \rho^i_\phi(z_i, s_i)  \right) \cdot \left( \oint \id z_i \ G(u-z_i) \sum_{s_i} \rho^i_\phi(z_i, s_i) \right) \Bigg] \label{B explicit}
\end{align}
with transition probability $\rho_\phi^i(z_i, s_i,\tau|z_{0,i}, s_{0,i})$. It is calculated in the next section.  

\subsection{Time dependent Fokker-Planck equation for the run-and-tumble particle at fixed \texorpdfstring{$\phi$}{ϕ}}\label{section time dependent Fokker Planck}
The time-dependent probability density $\rho_\phi(z, s, t)  = \prod_{i = 1}^N \rho_{\phi}^i(z_i, s_i,t)$ of the active particles at fixed profile $\phi(r)$ solves the time-dependent Fokker-Planck equation \cite{rtp_distribution, Bijnens_2021, Paoluzzi_2024, rtp_space_dependent_speed}
\begin{align}
     \frac{\partial \rho_\phi^i}{\partial t}(z_i, s_i, t) &= \mathcal{L}^\dagger_{z_i, s_i} \rho_\phi^i = -\frac{\partial}{\partial z_i} \left[\left(v_0 s_i + \mu f_{\phi}(z_i) \right) \rho_\phi^i(z_i, s_i,t) \right] + \alpha [\rho_\phi^i(z_i, - s_i,t) - \rho_\phi^i(z_i, s_i,t)] \label{time dependent fokker planck}
\end{align}
We drop the index $i$ in what follows since the equations are the same for all $i$. This partial differential equation (PDE) is supplied with appropriate initial and periodic boundary conditions
\begin{align*}
   \forall s,t: \qquad  \rho_\phi(0, s,t) &= \rho_\phi(L, s,t), \qquad \frac{\partial \rho_\phi}{\partial z}(0, s,t) = \frac{\partial \rho_\phi}{\partial z}(L, s,t), \qquad etc. \\
    \rho_\phi(z, s, 0)& = \delta_{s, s_0}\  \sum_{\ell = - \infty}^{\infty} \delta(z - z_0 - \ell L), \qquad  \delta_{s, s_0} = \frac{(1 + s s_0)}{2}
\end{align*}
By taking these initial conditions, the solution to \eqref{time dependent fokker planck} becomes the transition probability $ \rho_\phi(z, s, t) = \rho_\phi(z, s, t| z_0, s_0)$ appearing in \eqref{nu explicit}--\eqref{B explicit}. \\
We solve \eqref{time dependent fokker planck} by introducing the total particle (probability) density $p_\phi(z,t)$ and chirality $q_\phi(z,t)$
\begin{align*}
    p_\phi(z,t) &= \rho_\phi(z, s, t) + \rho_\phi(z, - s,t), \qquad s q_\phi(z,t) = \rho_\phi(z, s, t)- \rho_\phi(z, - s,t) \\
   \rho_\phi(z, s,t) &= \frac{p_\phi(z,t) + s q_\phi(z,t)}{2}
\end{align*}
which satisfy the coupled PDEs
\begin{align}
    \frac{\partial p_\phi}{\partial t}(z,t) & = - \frac{\partial }{\partial z} \left(v_0 q_\phi(z,t) + \mu f_\phi(z) \ p_\phi(z,t) \right) = -\frac{\partial J_\phi}{\partial z}(z,t) \label{coupled pde general 1} \\
    \frac{\partial q_\phi}{\partial t}(z,t) & = - 2 \alpha  \ q_\phi(z,t) - \frac{\partial}{\partial z} \left( v_0 p_\phi(z,t) + \mu f_\phi(z) \ q_\phi(z,t) \right) \nonumber
\end{align}
with initial and periodic boundary conditions
\begin{align*}
    p_\phi(z,0) & =    \sum_{\ell = - \infty}^{\infty} \delta(z - z_0 -  \ell L), \qquad  p_\phi(0,t) = p_\phi(L,t), \qquad \frac{\partial p_\phi}{\partial z}(0,t) = \frac{\partial p_\phi}{\partial t}(L,t), \qquad \textit{etc.} \\ 
    q_\phi(z,0) & =    s_0 \sum_{\ell = - \infty}^{\infty} \delta(z - z_0 -  \ell L), \qquad   q_\phi(0,t) = q_\phi(L,t), \qquad \frac{\partial q_\phi}{\partial z}(0,t) = \frac{\partial q_\phi}{\partial t}(L,t), \qquad \textit{etc.}
\end{align*}
In \eqref{coupled pde general 1}, we also introduced the current density
\begin{equation*}
    J_\phi(z,t) = v_0 q_\phi(z,t) + \mu f_\phi(z) \ p_\phi(z,t)
\end{equation*}

\subsection{Weak-coupling expansion}\label{section weak coupling}
As shown in \cite{dichotomousnoise}, these coupled equations cannot be solved for a general nonlinear force $f_\phi(z)$ since they are equivalent to an intricate integrodifferential equation in time for $p_\phi(z,t)$, with a kernel that involves the exponential of the differential operator $\partial_z$. In what follows, we take the weak coupling limit $\zeta_\phi \ll 1$ and expand
\begin{align}
    f_\phi(z) &= \zeta_\phi \tilde{f}_\phi(z), \qquad 
     \rho_\phi(z,s,t)    = \rho^0_\phi(z,s,t)  + \zeta_\phi \ \rho^1_\phi(z,s,t) + O(\zeta_\phi^2) \label{small coupling expansion} \\
     p_\phi(z,t) & = p^0_\phi(z,t) + \zeta_\phi \ p^1_\phi(z,t) + O(\zeta_\phi^2), \qquad     q_\phi(z,t)  = q^0_\phi(z,t) + \zeta_\phi \ q^1_\phi(z,t) + O(\zeta_\phi^2) \nonumber
\end{align}
Since $\nu, B$ are already of order $O(\zeta_\phi^2)$, we focus here only on the $O(\zeta_\phi^0)$ contribution in \eqref{small coupling expansion} which satisfy
and solve the PDEs perturbatively
\begin{align}
   O(\zeta_\phi^0) : \qquad  \frac{\partial p_\phi^0}{\partial t}(z,t) & = - v_0\frac{\partial q_\phi^0}{\partial z}(z,t), \qquad \frac{\partial q_\phi^0}{\partial t}(z,t) = - 2 \alpha q^0_\phi(z,t) - v_0 \frac{\partial p_\phi^0}{\partial z}(z,t) \label{leading order pde}
\end{align}
with initial and periodic boundary  conditions
\begin{align}
p_\phi^0(z,0) &= \sum_{\ell = - \infty}^{\infty} \delta(z - z_0 - \ell L), \qquad p_\phi^0(0, t) = p_\phi^0(L,t), \qquad \frac{\partial p_\phi^0}{\partial z}(0,t) = \frac{\partial p_\phi^0}{\partial z}(L,t), \qquad \textit{etc.}\label{conditions p q leading order}\\
q_\phi^0(z,0) &= s_0 \sum_{\ell = - \infty}^{\infty} \delta(z - z_0 - \ell L), \qquad q_\phi^0(0, t) = q_\phi^0(L,t), \qquad \frac{\partial q_\phi^0}{\partial z}(0,t) = \frac{\partial q_\phi^0}{\partial z}(L,t), \qquad \textit{etc.} \nonumber
\end{align}
Higher-order terms can easily be obtained following similar steps. \\
By differentiating, the leading order equations \eqref{leading order pde} can be converted into two separate telegrapher's equations for $p^0, q^0$,
\begin{align*}
 &\frac{\partial^2 p^0}{\partial t^2}(z,t) + 2 \alpha \frac{\partial p^0}{\partial t}(z,t) - v_0^2 \frac{\partial^2 p^0}{\partial z^2}(z,t) = 0, \qquad \frac{\partial^2 q^0}{\partial t^2}(z,t) + 2 \alpha \frac{\partial q^0}{\partial t}(z,t) - v_0^2 \frac{\partial^2 q^0}{\partial z^2}(z,t) = 0
\end{align*}
under the conditions \eqref{conditions p q leading order}. Because we differentiated, there are also the additional constraints
\begin{align*}
    \frac{\partial p^0}{\partial t}(z,0) & = - v_0 \frac{\partial q^0}{\partial z}(z,0), \qquad 
      \frac{\partial q^0}{\partial t}(z,0)  = - v_0 \frac{\partial p^0}{\partial z}(z,0) - 2 \alpha q^0(z,0) 
\end{align*}
These equations are solved by going to Fourier space, leading to
\begin{align}
    p^0(z,t) &= \frac{e^{- \alpha t}}{L} \sum_{n = - \infty}^{\infty} \left( \cosh(\Upsilon_n t) +  \frac{(\alpha -i s_0 v_0 2 \pi n/L)}{\Upsilon_n} \sinh(\Upsilon_n t) \right) e^{i2 \pi n(z-z_0)/L}  \label{telegraph solution delta source} \\
      q^0(z,t) &= \frac{e^{- \alpha t}}{L} \sum_{n = - \infty}^{\infty} \left( s_0 \cosh(\Upsilon_n t) -\frac{(\alpha s_0 + i v_0 2 \pi n/L)}{\Upsilon_n} \sinh(\Upsilon_n t) \right) e^{i2 \pi n(z-z_0)/L}   \label{telegraph solution delta source q} \\
     \rho^0(z, s,t) &= \frac{p^0(z,t) + s q^0(z,t)}{2} \label{telegraph solution rho}
    \end{align}
   with $ \Upsilon_n  = \sqrt{\alpha^2 -\frac{4 \pi^2n^2}{L^2} v_0^2}$. 
 Note that at late times
    \begin{equation*}
        \lim_{t \to \infty} p^0(z,t) = \frac{1}{L}
    \end{equation*}
as expected for the stationary distribution of independent, free (at leading order $O(\zeta_\phi^0)$) run-and-tumble particles on the ring. It can be shown that $p^0, q^0$ are real functions and converge in the distributional sense. \\
The solution $ \rho^0(z,s,t)$ in  \eqref{telegraph solution rho} is in agreement with the eigenfunction expansion of the Fokker-Planck equation  \cite{Kampen1992}, $\rho^0(z,s,t) = \sum_{\lambda_0} \rho^0_{\lambda_0}(z, s) \ e^{\lambda_0 t}$, with eigenvalues $\lambda_0$ and eigenfunctions $\rho^0_{\lambda_0}$
\begin{align*}
    \lambda_0 &= 0, \qquad \rho^0_0(z, s)  = \frac{1}{2L}\\
    \lambda_0 &= - 2 \alpha, \qquad \rho^0_{- 2 \alpha}(z, s) = \frac{s s_0}{2L} \\
    \lambda_0 &= - \alpha \pm \Upsilon_n, \qquad  \rho^0_{n, \pm}(z, s)  = \frac{1}{4L} \left( 1 + s s_0 \pm \frac{(\alpha (1 - s s_0) - i v_0 2 \pi n (s_0 + s)/L)}{\Upsilon_n} \right)  e^{i 2 \pi n(z-z_0)/L}
\end{align*}

\subsection{Calculating the covariances}
With the transition probability in hand, we compute \eqref{nu explicit}--\eqref{B explicit}. 
\subsubsection{Noise amplitude}
Following \eqref{Fourier space}, we expand the noise amplitude in a Fourier series
\begin{align}
 B(r,u) &= \sum_{n,m = - \infty}^{\infty} B_{n,m} \ e^{i 2 \pi n r/L} e^{i 2 \pi m u/L} \nonumber \\
     B_{n,m} 
     &  = k^2 G_n G_m \sum_i^N \int_0^{\infty} \id \tau \oint \id z_i  \ \id z_{0,i} \ e^{- i 2 \pi n z_i/L}  \ e^{- i 2 \pi m z_{i,0}/L} \label{B general delta potential} \\
     & \hspace{4 cm} \cdot  \Bigg[  \frac{P_\phi^i(z_i,\tau|z_{0,i}) \ p_\phi^i(z_{0,i}) + Q_\phi^{i}(z_i,\tau|z_{i,0}) \ q_\phi^i(z_{i,0})}{2}  -p^i_\phi(z_i) \ p^i_\phi(z_i)  \Bigg] \nonumber
\end{align}
where we have introduced the combinations
\begin{align}
    p_\phi^i(z_i) & = \rho_\phi^i(z_i, s_i) + \rho_\phi^i(z_i, - s_i), \qquad s_i q_\phi^i(z_i) = \rho_\phi^i(z_i, s_i) - \rho_\phi^i(z_i, - s_i) \nonumber \\
    p_\phi^i(z_i, \tau|z_{0,i},s_{0,i}) & =  \rho_\phi^i(z_i, s_i,\tau|z_{0,i}, s_{0,i}) + \rho_\phi^i(z_i, -s_i,\tau|z_{0,i}, s_{0,i}) \nonumber \\
    P_\phi^i(z_i,\tau|z_{0,i}) & =  p_\phi^i(z_i, \tau|z_{0,i},s_{0,i}) + p_\phi^i(z_i, \tau|z_{0,i},-s_{0,i}) \label{Big P}\\
    s_{i,0} Q_\phi^i(z_i,\tau| z_{0,i}) &= p_\phi^i(z_i, \tau|z_{0,i},s_{0,i}) - p_\phi^i(z_i, \tau|z_{0,i},-s_{0,i}) \label{Big Q}
\end{align}
Quantities without the time-parameter $\tau$ correspond to the stationary distributions. Equation \eqref{B general delta potential} is still the general (non-perturbative) result for the noise amplitude under a periodic potential. Focusing instead on the weak coupling limit $\zeta_\phi \ll 1$ and at leading order $O(\zeta_\phi^0)$ only, one finds with \eqref{telegraph solution delta source}--\eqref{telegraph solution delta source q},
\begin{align*}
    p_\phi^i(r) & = \frac{1}{L} + O(\zeta_\phi), \qquad q_\phi^i(u)  = O(\zeta_\phi)  \\
     P_\phi^i(r,\tau|u) & =\frac{2}{L} \left[1+ 2  e^{- \alpha \tau} \sum_{\ell = 1}^{\infty} \left( \cosh(\Upsilon_\ell \tau) +  \frac{\alpha}{\Upsilon_\ell} \sinh(\Upsilon_\ell \tau) \right) \cos\left(\frac{2 \pi \ell}{L}(r-u) \right) \right] + O(\zeta_\phi) \\
    Q_\phi^i(r,\tau|u) & = \frac{4 v_0}{L} e^{- \alpha \tau} \sum_{\ell = 1}^{\infty}   \frac{2 \pi \ell}{L\Upsilon_\ell} \sinh(\Upsilon_\ell \tau) \ \sin\left(\frac{2 \pi \ell}{L}(r-u) \right)  + O(\zeta_\phi)
\end{align*}
such that
\begin{align*}
    & \int_0^{\infty} \id \tau \oint \id z_i  \ \id z_{0,i} \ e^{- i 2 \pi n z_i/L}  \ e^{- i 2 \pi m z_{i,0}/L}  \Bigg[  \frac{P_\phi^i(z_i,\tau|z_{0,i}) \ p_\phi^i(z_{0,i}) + Q_\phi^{i}(z_i,\tau|z_{i,0}) \ q_\phi^i(z_{i,0})}{2}  -p^i_\phi(z_i) \ p^i_\phi(z_i)  \Bigg] \\
    & = \delta_{n,-m} \int_0^{\infty} \id \tau \   e^{- \alpha \tau}  \left( \cosh(\Upsilon_\ell \tau) +  \frac{\alpha}{\Upsilon_\ell} \sinh(\Upsilon_\ell \tau) \right)  + O(\zeta_\phi) \\
    & = \delta_{n,-m} (1 - \delta_{n,0}) \frac{L^2  \alpha }{2 \pi^2 v_0^2 n^2}
\end{align*}
With this, \eqref{B general delta potential} becomes
\begin{align}
    B_{n,m} &= \delta_{n,-m} (1 - \delta_{n,0})  k^2 \frac{ N L^2 \alpha}{2 \pi^2 v_0^2 n^2} |G_n|^2 + O(\zeta_\phi^3) \nonumber \\
      B (r,u)    &=  k^2 \frac{N L^2\alpha}{\pi^2 v_0^2} \sum_{n = 1}^{\infty} \frac{|G_n|^2}{n^2} \cos \left( \frac{2 \pi n}{L} (r-u) \right) + O(\zeta_\phi^3) \label{B(r,u) fourier}
\end{align}
 in agreement with Eq. (16) in the main text. There, it is explained that at leading order, $B$ is symmetric and positive-definite such that
the ``square root'' $\Gamma$ of $2 B(r,u)$ exists and satisfies
\begin{equation*}
    B(r,u)  = \frac{1}{2} \oint\ \id q \ \Gamma(r,q) \ \Gamma(u,q)
\end{equation*}
Since $B(r,u)$ depends only on $r-u$, we take the ansatz $\Gamma(r,u) = \Gamma(r-u)$ and expand in a Fourier series
yields
\begin{align}
\Gamma(x) & = \sum_{n = - \infty}^{\infty} \Gamma_n \ e^{i 2 \pi n x/L}, \qquad \Gamma_{-n} = \Gamma_n^* \label{fourier series gamma}\\
    \frac{1}{2} \oint\ \id q \ \Gamma(r,q) \ \Gamma(u,q) 
    & = \frac{L}{2} \Gamma_0^2 + L\sum_{n = 1}^\infty |\Gamma_n|^2 \cos\left(\frac{2 \pi n}{L}(r-u) \right)  \nonumber
\end{align}
Comparing this result with the Fourier series of $B$ in \eqref{B(r,u) fourier} implies
\begin{align*}
  \Gamma_n &=  (1 - \delta_{n,0}) \sqrt{N L \alpha } k \frac{|G_n|}{\pi v_0} \frac{e^{i \theta_n}}{|n|}, \qquad \Gamma(r,u)  =  2  k \frac{ \sqrt{N L \alpha}}{ \pi v_0} \sum_{n = 1}^{\infty} \frac{|G_n|}{n} \cos\left(\frac{2 \pi n}{L}(r-u) + \theta_n \right) \\
  \theta_{-n} &= -\theta_n + 2 \pi m_n, \qquad m_n \in \mathbb{Z}
\end{align*}
with (a priori) undetermined phase factor $\theta_n \in [0, 2 \pi)$ for $n \geq 1$. The ``square root'' of the noise amplitude is indeed not uniquely fixed, \cite{multivariatefokkerplanck, fokkerplanckbook}. However, if we require, similarly to $B$, that $\Gamma$ is positive definite and symmetric (\textit{i.e.} we take the principal, non-negative square root), one finds for $n \geq 1$
\begin{align*}
    & 0 \leq  \frac{1}{L^2} \oint \id r\oint \id u \ \chi(r) \  \Gamma(r,u) \ \chi(u)  =  2   k \frac{\sqrt{N \alpha}}{\pi v_0}  \sum_{n = 1}^{\infty} \frac{|G_n|}{n} \ \cos(\theta_n) \  |\chi_n|^2 \quad \Longrightarrow \quad \cos(\theta_n) \geq 0 \\
    & \Gamma(r,u) = \Gamma(u,r) \iff  \cos\left( \frac{2 \pi n (r-u)}{L} + \theta_n \right) = \cos\left( -\frac{2 \pi n (r-u)}{L} + \theta_n \right) \Longrightarrow  \theta_n = \pi m_n, \quad m_n \in \mathbb{Z}
\end{align*}
with unique solution $\theta_n = 0$ since $\theta_n \in [0, 2 \pi)$. Consequently
\begin{align*}
    \Gamma(r,u)  =  2  k \frac{ \sqrt{N L \alpha}}{ \pi v_0} \sum_{n = 1}^{\infty} \frac{|G_n|}{n} \cos\left(\frac{2 \pi n}{L}(r-u)  \right)
\end{align*}
which is Eq. (18) in the main text.

\subsubsection{Friction coefficient \texorpdfstring{$\nu$}{ν}}
To calculate the friction coefficient $\nu$ in \eqref{nu explicit}, we first note from \eqref{rho phi small coupling} that the derivative $  \frac{\delta \rho^i_{\phi}}{\delta \phi(u,t)}$ in Fourier space becomes
\begin{align*}
\phi_m(t) & = \frac{1}{L} \oint \id r \ \phi(r,t) \ e^{- 2 \pi i m r/L} \Longrightarrow \frac{\delta \phi_m(t)}{\delta \phi(u,t)} = \frac{1}{L} e^{- 2 \pi i m u/L} \\
    \frac{\delta \rho^i_{\phi}}{\delta \phi(u,t)}(z,s) &= \sum_{m = - \infty}^{\infty}   \frac{\delta \rho^i_{\phi}}{\delta \phi_m(t)}(z,s) \ \frac{\delta \phi_m(t)}{\delta \phi(u,t)}  \\
    &= - \zeta_\phi\frac{\mu}{2Lc}  \sum_{m = - \infty}^{\infty}  \left[\frac{2 \alpha }{ v_0}  - s \frac{2 \pi i m}{L}  \right] (1 - \delta_{m,0}) \ G_{-m} \  e^{ 2 \pi i m(z- u)/L}
\end{align*}
such that
\begin{align}
 \nu(r,u) &= \sum_{n,m = - \infty}^{\infty} \nu_{n,m} \ e^{i 2 \pi n r/L} e^{-i 2 \pi m u/L} \nonumber \\
   \nu_{n,m} 
   & =  k^2 \frac{G_n G_{-m} \ \mu Y }{2Lc} (1 - \delta_{m,0})  \sum_i^N   \int_0^{\infty} \id \tau \  \oint \id z_i \ \id z_{0,i} \ e^{-i 2 \pi  n z_i/L} e^{i2 \pi  m z_{i,0}/L} \label{nu n,m tussen} \\
   & \hspace{7 cm} \cdot \left(\frac{2\alpha}{v_0} P_\phi^i(z_i,\tau|z_{0,i}) - \frac{2 \pi i m}{L}  Q_\phi^i(z_i,\tau|z_{0,i}) \right) \nonumber
\end{align}
Using $P_\phi^i, Q_\phi^i$ from \eqref{Big P}--\eqref{Big Q}, we find
\begin{align*}
    & (1 - \delta_{m,0}) \int_0^\infty \id \tau\oint \id z_i \ \id z_{0,i} \ e^{- i2 \pi  n z_i/L} e^{i2 \pi  m z_{i,0}/L}  \left(\frac{2\alpha}{v_0} P_\phi^i(z_i,\tau|z_{0,i}) - \frac{2 \pi i m}{L}  Q_\phi^i(z_i,\tau|z_{0,i}) \right) \nonumber \\
    & = \delta_{n,m} (1 - \delta_{n,0}) \int_0^\infty \id \tau \  e^{- \alpha \tau}\left[  \frac{4 \alpha L}{v_0}  \left( \cosh(\Upsilon_n \tau) +  \frac{\alpha}{\Upsilon_n} \sinh(\Upsilon_n \tau) \right) -  \frac{8 \pi^2  n^2 v_0}{L}   \frac{\sinh(\Upsilon_n \tau)}{\Upsilon_n} \right] \\
    & = \delta_{n,m} (1 - \delta_{n,0}) \frac{2 L}{v_0} \left( \frac{ \alpha^2 L^2}{\pi^2 n^2 v_0^2}  -  1 \right)
\end{align*}
and \eqref{nu n,m tussen} becomes
\begin{align}
    \nu_{n,m} & =  k^2 \frac{|G_n|^2 \ N\mu Y }{v_0^2} (1 - \delta_{n,0}) \delta_{n,m} \left( \frac{\alpha^2 L^2}{ \pi^2 v_0^2 n^2} - 1\right) + O(\zeta_\phi^3) \nonumber  \\
    \nu(r,u) & = 2 k^2  \frac{N\mu Y }{v_0^2} \sum_{n = 1}^{\infty} |G_n|^2  \left( \frac{ \alpha^2 L^2}{\pi^2 v_0^2 n^2} - 1\right)\ \cos \left( \frac{2 \pi n}{L}  (r-u) \right) + O(\zeta_\phi^3)  \label{nu fourier expansion 2}
\end{align}
in agreement with Eq. (17) in the main text. \\
Note that a constant friction term $\nu_c \frac{\partial \phi}{\partial t}(r,t)$ in \eqref{modified klein gordon}, as appears in \textit{e.g.} \cite{brownianstring}, is not allowed since it requires $\nu(x)$ to be of the form
\begin{align*}
   \nu(x) = \nu_c \sum_{\ell = - \infty}^{\infty} \delta(x- \ell L) = \frac{\nu_c}{L} \sum_{n = - \infty}^{\infty} e^{i 2 \pi n x/L} = \frac{\nu_c}{L} + \frac{2\nu_c}{L}  \sum_{n = 1}^{\infty} \cos \left( \frac{2 \pi n}{L} x \right)
\end{align*} 
which is incompatible with \eqref{nu fourier expansion 2} since $\nu_0 = 0$,  $\nu_n$ is negative for large $n$ and it decays in $n$ due to the smoothness assumption of $G$.

\section{Langevin Klein-Gordon equation}\label{section Langevin klein gordon eq 2}
Up to  $O(\zeta_\phi^2)$, equation \eqref{modified klein gordon} reduces to
\begin{align}
   \frac{1}{c^2} \frac{\partial^2 \phi}{\partial t^2}(r,t) &  
     = \frac{\partial^2 \phi}{\partial r^2}(r,t)- M^2 \phi(r,t) - \oint \id u \  \mathcal{M}(r-u) \  \phi(u,t) - k \frac{N}{L} \oint \id z \ G(z)  \nonumber \\
     &\qquad -\oint \id u \  \nu (r-u) \frac{\partial \phi}{\partial t}(u,t)+\oint \id u \ \Gamma (r-u) \ \xi(u,t) + O(\zeta_\phi^3) \label{full klein gordon dirac delta}
\end{align}
with periodic boundary conditions $     \phi(0,t) = \phi(L,t),  \frac{\partial \phi}{\partial r}(0,t) = \frac{\partial \phi}{\partial r}(L,t)$, \textit{etc.} 
That gets most easily solved in Fourier space,
\begin{align}\label{eq general potential fourier space 2}
  \frac{\id^2 \phi_n}{\id t^2} &= - \omega_n^2 \phi_n(t) - k NG_0 c^2 \,\delta_{n,0} - \nu_{n} L c^2\frac{\id \phi_n}{\id t}(t) + \Gamma_{n} L  c^2 \xi_n(t) + O(\zeta_\phi^3) \\
   \omega_n &= c\sqrt{\frac{4  \pi^2 n^2}{L^2} + M_{\text{eff},n}^2}, \qquad  \nu_{n} =  (1 - \delta_{n,0}) k^2\frac{N Y |G_n|^2 \mu}{v_0^2} \left(\frac{L^2\alpha^2}{\pi^2 n^2c^2} - 1\right) \nonumber \\
  \Gamma_n &=  (1 - \delta_{n,0}) \sqrt{N L \alpha} k \frac{|G_n|}{\pi v_0} \frac{1}{|n|}, \qquad 
   \langle \xi_n(t) \ \xi_m^*(t') \rangle_\xi = \frac{\delta_{n,m}}{L} \delta(t-t') \nonumber
\end{align}
Assuming mode $n$ to be underdamped, $\frac{\nu_n L c^2}{2 \omega_n} < 1$, it oscillates with a reduced frequency $\Omega_n$ (compared to the undamped case) given by, to order $\zeta_\phi^2$,
\begin{equation} \label{def omega_n 2}
    \Omega_n^2 = \omega_n^2 - \frac{\nu_{n}^2 L^2}{4}  c^4 = c^2 \left( \frac{4  \pi^2 n^2}{L^2} + M^2 - (1 - \delta_{n,0}) k^2 \frac{N}{L} \,Y |LG_n|^2\,\frac{ 2 \alpha \mu}{v_0^2} \right) 
\end{equation}
Then, the solution to \eqref{eq general potential fourier space 2} is, \cite{brownianstring},
\begin{align}
    \phi_0(t) & =   \phi_0(0)  \cos\left(M vt\right)  +  \frac{ \phi_0'(0)}{Mv} \sin\left( M c t\right) -k \frac{N G_0}{ M^2} (1 - \cos(M c t)) \label{solution phi_0(t)} \\
      \phi_n(t) & = e^{- \frac{\nu_n L}{2} c^2 t} \left[ \phi_n(0) \cos\left(\Omega_n t \right) + \frac{\nu_n L c^2 \phi_n(0) + 2 \phi'_n(0)}{2 \Omega_n} \sin\left(\Omega_n t \right) \right] \label{hom part} \\
      & \qquad + c^2  \Gamma_{n} L \int_0^{t} \id \tau \  e^{- \frac{\nu_{n} L}{2} c^2(t-\tau)} \frac{\sin\left(\Omega_n (t-\tau) \right)}{\Omega_n} \xi_n(\tau) , \quad n \neq 0 \nonumber 
\end{align}

\subsection{Variance}\label{vari}
Using the properties of the noise in Eq. (8) of the main text, the modes $n \neq 0$ satisfy
\begin{align}
    \langle \phi_n(t) \rangle_\xi &=  e^{- \frac{\nu_n L}{2} c^2 t} \left[ \phi_n(0) \cos\left(\Omega_n t \right) + \frac{\nu_n L c^2 \phi_n(0) + 2 \phi'_n(0)}{2 \Omega_n} \sin\left(\Omega_n t \right) \right] \label{average phi n supp}\\
    \text{Var}\left(\phi_n(t) \right)_\xi & = \langle \phi_n(t) \ \phi_n^*(t) \rangle_\xi - \langle \phi_n(t) \rangle_\xi \ \langle \phi_n^*(t) \rangle_\xi = \langle \phi_n(t) \ \phi_{-n}(t) \rangle_\xi - \langle \phi_n(t) \rangle_\xi \ \langle \phi_{-n}(t) \rangle_\xi \nonumber \\
    &=  \frac{c^4  |\Gamma_{n}|^2 L}{\Omega_n^2}\int_0^{t} \id \tau  \  e^{- \nu_n L c^2(t-\tau)} \sin^2\left(\Omega_n (t-\tau) \right) \nonumber \\
      & =  \frac{ |\Gamma_{n}|^2 c^2}{2 \Omega_n^2 \nu_{n}  \omega_n^2}  \Bigg[ \Omega_n^2  - \omega_n^2 e^{-\nu_{n} Lv^2 t} + \frac{\nu_{n} c^2 L}{2} e^{-\nu_{n} Lv^2 t}\left(\frac{\nu_{n} c^2L}{2} \cos\left(2 \Omega_n t\right) - \Omega_n \sin\left(2 \Omega_n t\right) \right)  \Bigg]  \label{variance phi_n}
 \end{align}
We thus recognize two possibilities as $t \to \infty$ for $\text{Var}\left(\phi_n(t) \right)_\xi$ (with $n \neq 0$):
\begin{itemize}
    \item If $\nu_{n} > 0$ (when in equilibrium or possibly for small $n$), then most terms decay exponentially, leading to a constant variance
    \begin{align}
       \lim_{t \to \infty} \text{Var}\left(\phi_n(t) \right)_\xi = \frac{|\Gamma_{n}|^2 c^2}{2 \nu_{n} \omega_n^2} =\frac{\alpha  L}{2 \pi^2 n^2  \mu Y \left(\frac{4 \pi^2}{L^2} n^2 + M_{\text{eff},n}^2 \right)} \left( \frac{L^2\alpha^2}{\pi^2 n^2 v_0^2} -1\right)^{-1} \label{fluctuations phi n at late times supp}
    \end{align}
and the membrane is fluctuating. In equilibrium, this becomes
     \begin{align*}
        \lim_{t \to \infty}  \text{Var}\left(\phi_n(t) \right)_\xi^{\text{eq}} &=\frac{k_B T}{2L Y} \left(\frac{4 \pi^2 n^2}{L^2} + M_{\text{eff},n,\text{eq}}^2 \right)^{-1} \\
        &=\frac{k_B T}{2 L Y}  \left(\frac{4 \pi^2 n^2}{L^2} + M^2 - k^2 \frac{N L  Y |G_n|^2}{k_B T}\right)^{-1} 
    \end{align*}
    which decays with $n$ and $L$. 
        \item  If $\nu_{n} < 0$ (for large $n$), then most terms grow exponentially, indicating an instability. \
\end{itemize}

\subsection{Inverse Landau-like damping}
The homogeneous part of the solution \eqref{hom part} has the form $\phi_{h,n} = A_{n} e^{i W_n^{+} t} + B_{n} e^{i W_n^{-} t}$ where $A_n, B_n$ encode the initial conditions
\begin{align*}
    A_{n}  = -\frac{i \phi'_{h,n}(0) + W^{-}_{n} \phi_{h,n}(0)}{W^{+}_{n} - W^{-}_{n}}, \qquad B_{n} = \frac{ W^{+}_{n} \phi_{h,n}(0) + i \phi_{h,n}'(0)}{W^{+}_{n} - W^{-}_{n}}
\end{align*}
and the $W_n^{\pm}$ satisfy the dispersion relations
\begin{align*}
    &W_n^{\pm} = i \frac{\nu_{n}L c^2}{2} \pm \Omega_n = i  (1 - \delta_{n,0}) k^2 \frac{N Y L \mu |G_n|^2 c^2}{2c^2} \left( \frac{L^2\alpha^2}{\pi^2 n^2 v_0^2}  - 1 \right) \\
    &  \qquad \pm  c \sqrt{\frac{4 \pi^2 n^2}{L^2} + M^2  - \zeta_\phi^2 (1-\delta _{0,n}) \frac{ N L 2\alpha \mu |G_n|^2}{Y v_0^2} \left(1  + \zeta_\phi^2 \frac{N L \mu |G_n|^2 c^2}{8Y\alpha v_0^2} (1-\delta_{0,n}) \left(1-\frac{ L^2\alpha^2}{\pi^2 n^2 v_0^2}\right)^2  \right)} \\
     & \Longrightarrow W_0^{\pm} = \pm  Mv,
        \qquad  W_n^{\pm} \sim - i \frac{N Y L \mu |G_n|^2 c^2}{2c^2} \pm  \frac{2 \pi n}{L} \ c
    \quad \text{ for } \ n \gg 1
\end{align*}
where the imaginary part induces a positive or negative friction. Hence, the $n = 0$ mode oscillates without friction, the $n \sim  O(1)$ modes experience positive or negative friction (depending on the ratio $\frac{L \alpha}{\pi v_0}$), while large modes $n \gg 1$ experience a negative friction, vanishing with $n \to \infty$.

\section{Numerical simulations}
In this section, we confirm the emergence of negative friction in the dynamics \eqref{original klein gordon equation supp}--\eqref{original z dynamics} with numerical simulations. Going to Fourier space yields
\begin{align}
    &\frac{1}{c^2} \frac{\id^2 \phi_n}{\id t^2} + \left( \frac{4 \pi^2 n^2}{L^2} + M^2 \right) \phi_n = - k G_n\sum_{j = 1}^N  \ e^{- \frac{2 \pi i n}{L} z_j(t)} \\
    & \frac{\id z_j}{\id t} =  v_0 s_j(t)+ 4 \pi \zeta_\phi \mu \sum_{n = 1}^{\infty} n \ \text{Im}\{\phi_n \ G_{n}^* \ e^{\frac{2 \pi i n}{L} z_j(t)}\} \label{z dynamics with all modes}
\end{align}
where $\phi_{-n} = \phi_n^*$ and $G_{-n} = G_n^*$ since $\phi, G$ are real functions, implying we only need to focus on $n \geq 0$. Equation \eqref{z dynamics with all modes} contains all modes $\phi_n$ unless $G_n$ vanishes for  $ n > n_0, \quad n_0 \in \mathbb{N}$. To demonstrate via simulations the occurrence of negative friction, we consider the case with
\begin{align*}
    G(x) = g \left(1+ \cos\left( \frac{2 \pi }{L} x\right) \right) \Longrightarrow G_0 = g, \qquad  G_1 =  \frac{g}{2}, \qquad G_{n} = 0 \ \text{ for } \  n > 1
\end{align*}
resulting in
\begin{align}
& \frac{1}{c^2} \frac{\id^2 \phi_0}{\id t^2} + M^2 \phi_0 = - k  Ng \label{dynamics phi 0}, \qquad \frac{1}{c^2} \frac{\id^2 \phi_n}{\id t^2} + \left( \frac{4 \pi^2 n^2}{L^2} + M^2 \right) \phi_n = 0 \quad \text{ for $n > 1$} \\
 &\frac{1}{c^2} \frac{\id^2 \phi_1}{\id t^2} + \left( \frac{4 \pi^2}{L^2} + M^2 \right) \phi_1 =  -k\frac{g}{2} \sum_{j = 1}^N  \ e^{- \frac{2 \pi i}{L} z_j(t)} \label{full dynamics phi1} \\
    & \frac{\id z_j}{\id t} =  v_0 s_j(t)+ 2 \pi \zeta_\phi \mu g \ \Im\{\phi_1  \ e^{\frac{2 \pi i }{L} z_j(t)}\} \label{z coupled to phi1}
\end{align}
with imaginary part denoted by $\Im\{\cdot\}$. Equation \eqref{dynamics phi 0} for $\phi_0$ is the same as \eqref{eq general potential fourier space 2} with solution \eqref{solution phi_0(t)}. Moreover, there is only a coupling between $z_j$ and $\phi_1$ while the other modes $\phi_n$ for $n > 1$ are free. In the limit $c \ll v_0$, the coupled equations \eqref{full dynamics phi1}--\eqref{z coupled to phi1} should be compared to the reduced dynamics \eqref{eq general potential fourier space 2} for $n = 1$,
\begin{align}
    \left(\frac{\id^2}{\id t^2} + \nu_1 L c^2 \frac{\id}{\id t} + \omega_1^2 \right) \phi_1(t) &= \Gamma_{1} L c^2  \xi_1(t) \label{eq cosine potential fourier space} 
\end{align}
The initial conditions $\phi_n(0) = \phi_n'(0) = 0$ for $n > 1$ yield $\phi_n(t) = 0$ for $n > 1$ such that the field solution $\phi(r,t)$ in real space reduces to
\begin{align*}
    \phi(r,t) &= \phi_0(t) + \phi_1(t) \ e^{2 \pi i r/L} + \phi_{1}^*(t) \ e^{-2 \pi i r/L} \\
    & = \phi_0(t) + 2 \ \Re\{ \phi_1(t) \ e^{2 \pi i r/L} \} \\
    & = \phi_0(t) + 2 \ \Re\{ \phi_1(t) \} \ \cos\left(\frac{2 \pi }{L} r \right) -  2 \ \Im\{ \phi_1(t) \} \ \sin\left(\frac{2 \pi }{L} r \right)
\end{align*}

\subsection{Numerical implementation}
We simulate the equations \eqref{full dynamics phi1}--\eqref{z coupled to phi1}, which we write in the form
\begin{equation}\label{simulation eq}
    \begin{aligned}
        \frac{\id \phi_1}{\id t}(t) &= \psi_1(t), \qquad      \frac{\id \psi_1}{\id t}(t) =  F_\phi(\phi_1, \zvec{z})\\
        \frac{\id z_j}{\id t}(t) &= v_0 s_j(t) + F_z(\phi_1, z_j) , \quad j=1,\cdots,N.
    \end{aligned}
\end{equation}
with forces $F_\phi, F_z$
\begin{align*}
    F_\phi(\phi_1, \zvec{z}) & = - \left( \frac{4 \pi^2 n^2}{L^2} + M^2 \right)  c^2 \phi_1   -k\frac{g c^2}{2} \sum_{j = 1}^N  \ e^{- \frac{2 \pi i}{L} z_j(t)} \\
    F_z(\phi_1, z_j) & =  2 \pi\zeta_\phi \mu g \ \Im\{\phi_1  \ e^{\frac{2 \pi i }{L} z_j(t)}\}
\end{align*} 
The full state of the system is described by the variables $\phi_1, \psi_1, \vec{z},\vec{s}$ where the propulsion direction of each RTP, $s_j$, flips at random times obtained from an exponential distribution with flipping rate $\alpha$. As initialisation, we use fixed initial conditions for the field mode $\phi_1(0), \psi_1(0)$ while the positions $\vec{z}(0)$ and the spin direction $\vec{s}(0)$ of the RTPs are randomly chosen. Moreover, we generate $N$ independent exponentially distributed random numbers $\tau_i$, representing the (first) spin flip time $s_i$. \\ Next, we discretize the time evolution \eqref{simulation eq} in $N_{\text{steps}} \in \mathbb{N}$ timesteps of size $\id t$ such that $t_{m + 1} = t_m + \id t$ for $m = 0,..., N_{\text{steps}} -1$. At the beginning of the time step $t_m \to t_{m+1}$, we decrease the flip time $\tau_j^m \to \tau_j^{m+1} = \tau_j^m-\id t$ if $\tau_j^m> \id t$ as the flip event comes closer. Alternatively, if $\tau_j^m< \id t$, the spin $s_j$ will flip during the step $t_m \to t_{m+1}$ such that we change its sign $s_j \to - s_j$, and generate a new flip time $\tau_i$ according to the exponential distribution. Following a second-order Runge-Kutta method, \cite{rungekutta} and supposing that at time $t_m$ we have the values $\phi_1^m, \psi_1^m, z_j^m, s_j^m$, the formula for the updated quantities $\phi_1^{m+1}, \psi_1^{m+1}, z_j^{m+1}$ at time $t_{m + 1} $ is as follows for the mode $\phi_1, \psi_1$ and the RTPs that do not flip
    \begin{align*}
        &\Delta_1 z_j = \left(F_z(\phi_1^m,z^m_j)+v_0s_j^m\right)\ \id t\\
        &\Delta_1\phi_1 = \psi_1^m \ \id t, \qquad \Delta_1 \psi_1 =F_\phi(\phi^m,\zvec{z}^m) \ \id t\\
        &\Delta_2 z_j = \left(F_z(\phi_1^m+\Delta_1\phi_1,z^m_j+\Delta_1 z_j)+v_0 s_j^m \right) \ \id t\\
        &\Delta_2 \phi_1 = \left(\psi_1^m + \Delta_1 \psi_1 \right) \ \id t \qquad \Delta_2\psi_1 = F_\phi(\phi_1^m+\Delta_1\phi_1,\zvec{z}^m+\Delta_1 \zvec{z}) \ \id t\\
        &z_j^{m+1}=z_j^m+\frac{\Delta_1z_j+\Delta_2z_j}{2},\\
        &\phi_1^{m+1}=\phi_1^m+\frac{\Delta_1\phi_1+\Delta_2\phi_1}{2}, \qquad \psi_1^{m+1}=\psi_1^m + \frac{\Delta_1\psi_1+\Delta_2\psi_1}{2}
    \end{align*}
    while for RTPs that do flip, the rules for $\Delta_1z_j$ and $\Delta_2z_j$ in the above change to 
\begin{align*}
    &\Delta_1 z_j = F_z(\phi_1^m,z^m_j) \ \id t+v_0 s_j^m \ (2\tau_j^m-\id t),\\
    &\Delta_2 z_j = F_z(\phi_1^m+\Delta_1\phi_1,z^m_j+\Delta_1 z_j) \ \id t+v_0 s_j^m(2\tau_j^m-\id t),
\end{align*}
where $\tau_j^m <\id t$ is the time of flipping. After all, for $t_m < t < t_m + \tau$ the spin has value $s_i^m$, while for the remaning time $t_m + \tau < t < t_{m} + \id t$ the spin has value $-s_i^m$, leading to an overal change $s_i^m \tau_j^m  - s_i^m (\id t - \tau_j^m) = s_i^m (2 \tau_j^m - \id t)$. The above evolution rule is accurate up to the second order of $\id t$. 

\subsection{Figures}

In the Figures below, we indicate the average and standard deviation over noise/spin realizations with $\langle \cdot \rangle_s$, resp. $\Delta \left( \cdot \right)_s$. The code is publicly available; see \cite{numericalsimulation}. To clearly observe the induced (positive or negative) friction effect, we take the parameter values in Table \ref{table parameter values}.
\begin{table}[H]
\centering
\caption{Simulation parameters used in this study.}
\begin{tabular}{llll}
\toprule
\textbf{Parameter} & \textbf{Symbol} & \textbf{Value} & \textbf{Unit} \\
\midrule
Ring length              & $L$             & 25$\pi$            & m \\
Wave velocity   & $c$        & 1           & m s$^{-1}$ \\
Young modulus         & $Y$      & 10          &  N \\
Spring constant (per unit length)         & $\kappa$      & 0.1          & N m$^{-2}$ \\
Initial mode value       & $\phi_n(0)$             & 1            & m \\
Initial mode speed       & $\phi'_n(0)$             & 2            & m s$^{-1}$\\
Coupling constant       & $\zeta_\phi$             & 0.1            & - \\
Force amplitude (per unit length) & $g$             & 1            & N m$^{-1}$ \\
Mobility       & $\mu$             & 1            & m s$^{-1}$ N$^{-1}$ \\
RTP speed       & $v_0$             & 20            & m s$^{-1}$ \\
Number of colloids       & $N$             & 20            & - \\
\bottomrule
\label{table parameter values}
\end{tabular}
\end{table}
For these values, the critical tumbling rate $\alpha_c = \pi v_0/L = 0.8$. In what follows, we study the negative ($\alpha < \alpha_c$) and positive ($\alpha > \alpha_c$) friction regimes
\subsubsection{Negative friction}
For $\alpha = 0.1 < \alpha_c$, we obtain
\begin{figure}[H]
\centering
  \includegraphics[width=0.6\linewidth]{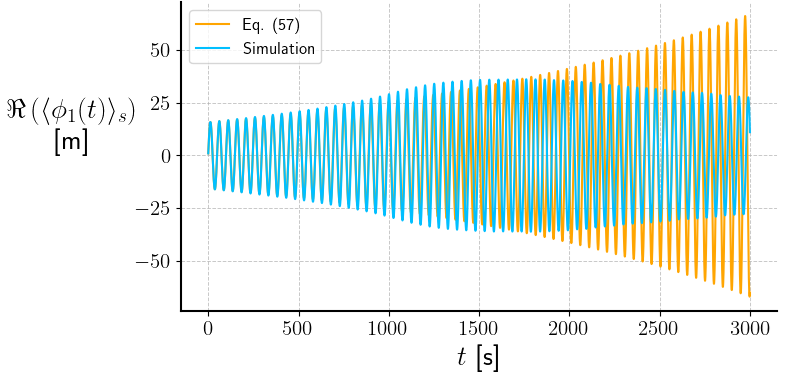}
\caption{Average of $\Re\{\phi_1(t)\}$ over 100 noise/spin realisations versus time. At short times, $t < 1000$ s, the simulation result and Eq. \eqref{average phi n supp} overlap almost exactly and exhibit exponential growth due to the negative friction effect. At later times $t > 1500 $ s, the simulation amplitude saturates, while \eqref{average phi n supp} keeps growing.}
\label{average phi1 rtp negative friction supp}
\end{figure}

\begin{figure}[H]
    \centering
    \includegraphics[width=0.6\linewidth]{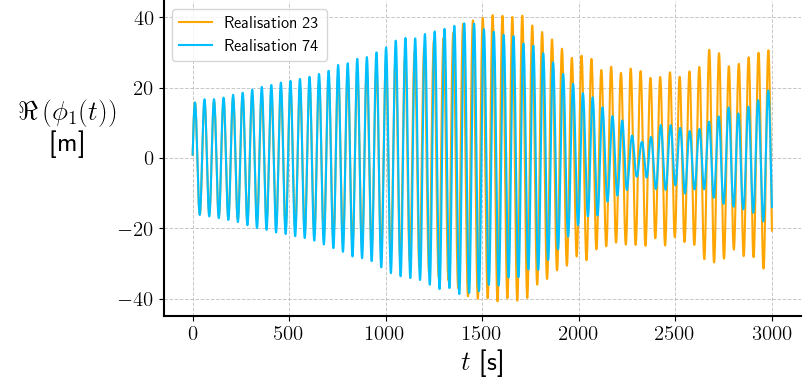}
    \caption{Individual $\Re\{\phi_1\}$ realisations (no average). We clearly see the negative friction effect due to the initial growth, but eventually the amplitude saturates, after which it shows pulsations of growth and decay.}
    \label{individual trials rtp negative friction}
\end{figure}
\begin{figure}[H]
    \centering
    \includegraphics[width=0.6\linewidth]{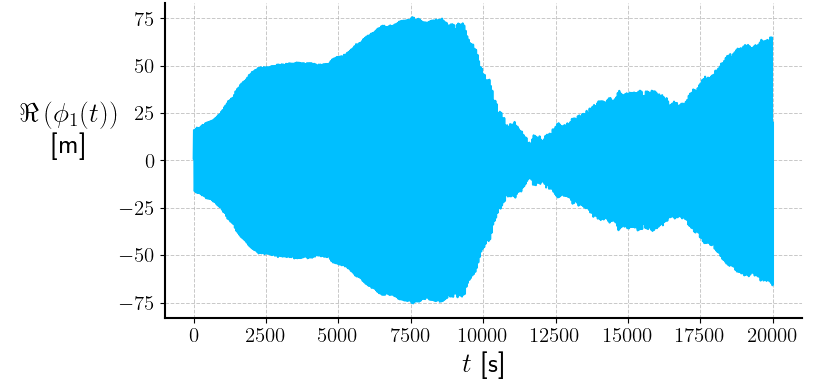}
    \caption{Individual $\Re\{\phi_1\}$ trajectory (no average) {\it vs} time, simulated over a long time interval. On this timescale, one cannot distinguish the individual oscillations, but we clearly see the negative friction effect due to the initial growth. Eventually, the amplitude saturates and shows pulsations of growth and decay.}
    \label{individual trials rtp negative friction long time supp}
\end{figure}

Taking fewer noise realizations but simulating for a longer time gives Fig.~\ref{individual trials rtp negative friction long time supp}.

\subsubsection{Positive friction}
We can also get positive friction by increasing the tumbling rate, \textit{e.g.} for $\alpha = 1 > \alpha_c$, we find 
\begin{figure}[H]
\centering
  \includegraphics[width=0.6\linewidth]{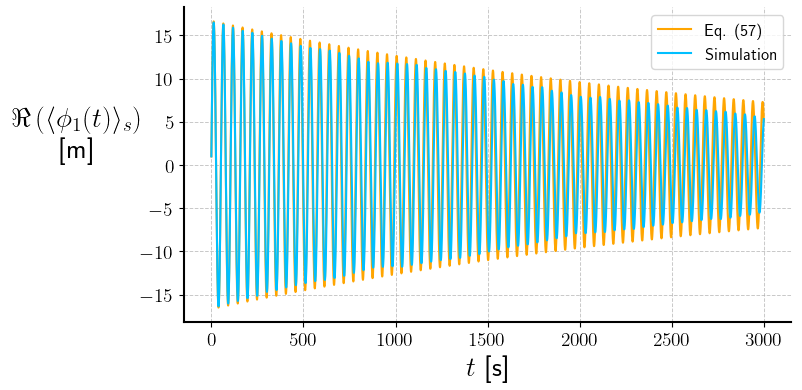}
\caption{Average of $\Re\{\phi_1(t)\}$ over 100 noise/spin realisations versus time. The simulation result and Eq. \eqref{average phi n supp} overlap and exhibit exponential decay over the entire time range due to the positive friction effect.}
\label{average phi1 rtp positive friction}
\end{figure}

\begin{figure}[H]
    \centering
\includegraphics[width=0.6\linewidth]{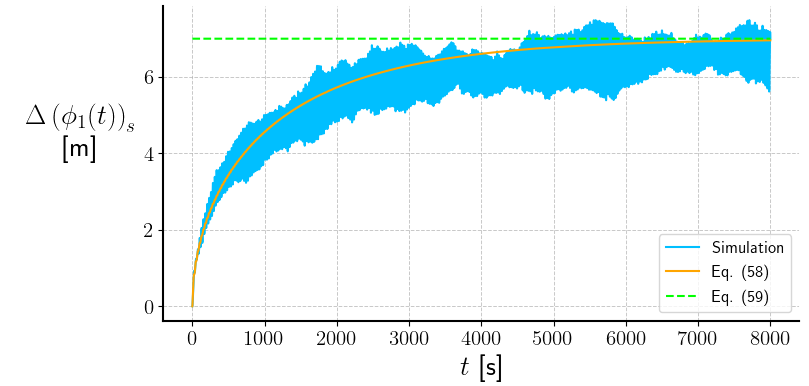}
    \caption{Standard deviation of $\phi_1$ over 100 noise/spin realisations versus time. The simulation result and Eq. \eqref{variance phi_n} overlap almost exactly and converge to the expected result \eqref{fluctuations phi n at late times supp} at late times.}
    \label{variance phi 1 rtp positive friction}
\end{figure}

\end{document}